\documentclass[reprint,amsfonts,showpacs,pas,aip,english,superscriptaddress,float]{revtex4-1}

\usepackage{graphicx}
\usepackage{overpic}
\usepackage{times}
\usepackage{CJK}
\usepackage{amsmath}

\draft

\begin{document}
\title{Effective suppression of parametric instabilities with decoupled broadband lasers in plasma}

\author{Yao Zhao}
\affiliation{Key Laboratory for Laser Plasmas (MoE), School of Physics and Astronomy, Shanghai Jiao Tong University, Shanghai 200240, China}
\affiliation{Collaborative Innovation Center of IFSA (CICIFSA), Shanghai Jiao Tong University, Shanghai 200240, China}

\author{Suming Weng}
\affiliation{Key Laboratory for Laser Plasmas (MoE), School of Physics and Astronomy, Shanghai Jiao Tong University, Shanghai 200240, China}
\affiliation{Collaborative Innovation Center of IFSA (CICIFSA), Shanghai Jiao Tong University, Shanghai 200240, China}

\author{Min Chen}
\affiliation{Key Laboratory for Laser Plasmas (MoE), School of Physics and Astronomy, Shanghai Jiao Tong University, Shanghai 200240, China}
\affiliation{Collaborative Innovation Center of IFSA (CICIFSA), Shanghai Jiao Tong University, Shanghai 200240, China}

\author{Jun Zheng}
\affiliation{Key Laboratory for Laser Plasmas (MoE), School of Physics and Astronomy, Shanghai Jiao Tong University, Shanghai 200240, China}
\affiliation{Collaborative Innovation Center of IFSA (CICIFSA), Shanghai Jiao Tong University, Shanghai 200240, China}

\author{Hongbin Zhuo}
\affiliation{College of Science, National University of Defense Technology, Changsha 410073, China }
\affiliation{Collaborative Innovation Center of IFSA (CICIFSA), Shanghai Jiao Tong University, Shanghai 200240, China}

\author{Chuang Ren}
\affiliation{Department of Mechanical Engineering and Laboratory for Laser Energetics, University of Rochester, Rochester, New York 14627, USA}
\affiliation{Department of Physics and Astronomy, University of Rochester, Rochester, New York 14627, USA}

\author{Zhengming Sheng}
\email[]{Email: zmsheng@sjtu.edu.cn}
\affiliation{Key Laboratory for Laser Plasmas (MoE), School of Physics and Astronomy, Shanghai Jiao Tong University, Shanghai 200240, China}
\affiliation{Collaborative Innovation Center of IFSA (CICIFSA), Shanghai Jiao Tong University, Shanghai 200240, China}
\affiliation{SUPA, Department of Physics, University of Strathclyde, Glasgow G4 0NG, UK}
\affiliation{Tsung-Dao Lee Institute, Shanghai Jiao Tong University, Shanghai 200240, China}

\author{Jie Zhang}
\affiliation{Key Laboratory for Laser Plasmas (MoE), School of Physics and Astronomy, Shanghai Jiao Tong University, Shanghai 200240, China}
\affiliation{Collaborative Innovation Center of IFSA (CICIFSA), Shanghai Jiao Tong University, Shanghai 200240, China}

\date{\today}

\begin{abstract}
A theoretical analysis for the stimulated Raman scattering (SRS) instability driven by two laser beams with certain frequency difference is presented. It is found that strong coupling and enhanced SRS take place only when the unstable regions corresponding respectively to the two beams are overlapped in the wavenumber space. Hence a threshold of the beam frequency difference for their decoupling is found as a function of their intensity and plasma density. Based upon this, a strategy to suppress the SRS instability with decoupled broadband lasers (DBLs) is proposed. A DBL can be composed of tens or even hundreds of beamlets, where the beamlets are distributed uniformly in a broad spectrum range such as over 10\% of the central frequency. Decoupling among the beamlets is found due to the limited beamlet energy and suitable frequency difference between neighboring beamlets. Particle-in-cell simulations demonstrate that SRS can be almost completely suppressed with DBLs at the laser intensity $\sim10^{15}$ W/cm$^2$. Moreover, stimulated Brillouin scattering (SBS) will be suppressed simultaneously with DBLs as long as SRS is suppressed. DBLs can be attractive for driving inertial confined fusion.
\end{abstract}

\maketitle

\section{Introduction}

Campaigns to achieve ignition on National Ignition Facility (NIF) yielded significant insights of inertial confinement fusion (ICF) \cite{Betti2016Inertial,Lindl2014Review}, including reaching the milestone of fuel gain exceeding unity \cite{Hurricane2014Fuel}. Meanwhile, a few critical challenges to further enhance the laser-target energy coupling efficiency have been revealed. Currently there are no clear paths to ignition on NIF or similar-sized facility. Exploring alternative approaches is necessary. Laser plasma instabilities (LPI) are among the major obstacles to both direct- and indirect-drive schemes, causing asymmetric \cite{Town2014Dynamic,Moody2014Early} and insufficient drive \cite{Moody2014Progress,Igumenshchev2012Crossed} and preheating \cite{Regan2010Suprathermal,Smalyuk2008Role,Sangster2008High}. A few ideas have been proposed to suppress LPI by use of various beam smoothing techniques \cite{skupsky1989improved,lehmberg1983use,froula2010experimental,moody2001backscatter}, temporal profile shaping \cite{B2014Control}, laser beams with broadband width \cite{thomson1974effects,Eimerl}, and enhanced plasma damping \cite{turner1985evidence,craxton2015direct}, etc. However, it is not possible to suppress LPI completely.

In this work, we present a theory, backed by particle-in-cell (PIC) simulations, that a new type of lasers called decoupled broadband lasers (DBLs) can completely suppress stimulated Raman scattering (SRS), a major concern to both direct- and indirect-drive ICF \cite{InterplayStrozzi}. A DBL is composed of many beamlets, which may have different frequencies among beamlets within certain range. A related idea is the Coherent Amplification Network (CAN) \cite{mourou2013future}. Different from the CAN scheme, here the required laser power of DBLs for fusion application is much lower and there is no requirement for the phase lock between DBL beamlets, which is called incoherent combination \cite{andrusyak2009beam,hamilton2004high,farmer1999incoherent,benedetti2014plasma}. More recently, a concept of broadband laser driver called StarDriver was proposed for ICF application to control both hydrodynamic and laser-plasma instabilities \cite{eimerl2016stardriver}, where a laser driver is consisted of many beamlets at an aperture. Physically, it is not clear so far whether and how a broadband laser driver may suppress the laser plasma instabilities. In this work, we will clarify the mechanism and conditions on DBLs for almost complete suppression of the SRS instability based upon theoretical and numerical studies.

\section{Model of two beam coupling and decoupling}

We first introduce a model for DBLs. The temporal part of such light can be written as
\begin{equation} \label{1}
a_\textsc{dbl}=\sum_{i=1}^Na_{i}\cos(\omega_it+\phi_i),
\end{equation}
where $a_i$ is the normalized amplitude of the $i$-th beamlet with a carrier frequency $\omega_i$, $\phi_i$ is a random phase between $[-\pi,\pi]$, and $N$ is the number of beamlets typically around a few hundreds. The beamlets are nearly uniformly distributed in the total frequency spectrum bandwidth $\Delta\omega_0$, as shown schematically in Fig. 1. Here the amplitude $a_i$ is related to the light intensity $I_i$ given by $a_i=\sqrt{I_i(\mathrm{W}/\mathrm{cm}^2)[\lambda_i(\mu\mathrm{m})]^2/1.37\times10^{18}}$. Before the study of a DBL propagation in plasma, we first consider the coupling of two light components $(k_i,\omega_i)$ with $i=1$ or 2, where $k_i$ and $\omega_i$ are the laser wavenumber and frequency, respectively.

\begin{figure}
    \begin{tabular}{lc}
        \begin{overpic}[width=0.5\textwidth]{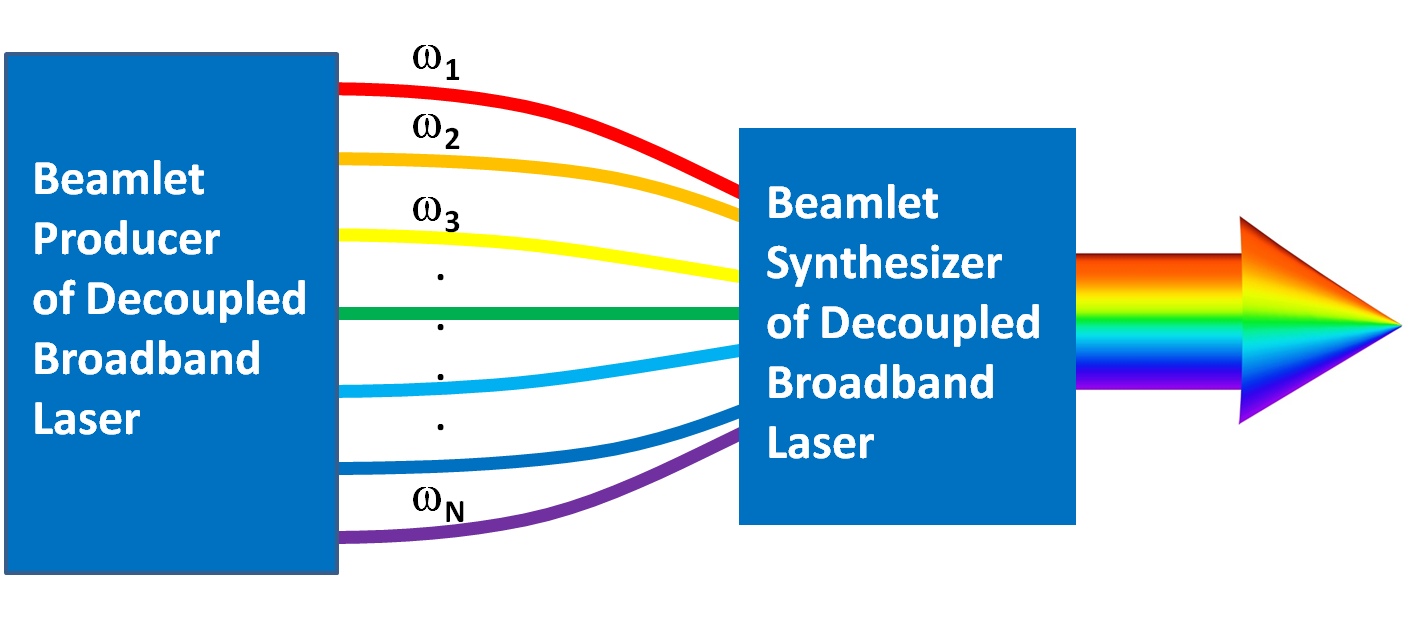}
        \end{overpic}
    \end{tabular}
\caption{ A decoupled broadband laser beam is composed of many beamlets such as 100 beamlets with a frequency difference larger than 0.1\% between every two adjacent-frequency beamlets.
    }
\end{figure}

Let $\omega_1=\omega_0+\delta\omega/2$ and $\omega_2=\omega_0-\delta\omega/2$, where $\omega_0$ and $k_0$ are the center frequency and center wavenumber, respectively, and $\delta\omega$ is the frequency difference between them. Under the condition $\delta\omega\lesssim10^{-2}\omega_0$, we can write $k_1=k_0+\delta k/2$ and $k_2=k_0-\delta k/2$, where $\delta k=k_0\omega_0\delta\omega/(\omega_0^2-\omega_{pe}^2)$, and $\omega_{pe}$ is the electron plasma frequency. The coupled fluid equations for SRS backscattering are
\begin{equation} \label{2}
\left(\frac{\partial^2}{\partial t^2}-c^2\nabla^2+\omega_{pe}^2\right)\widetilde{A}=-4\pi ec^2\tilde{n}_ea_{DBL},
\end{equation}
\begin{equation} \label{3}
\left(\frac{\partial^2}{\partial t^2}-3v_{th}^2\nabla^2+\omega_{pe}^2\right)\tilde{n}_e=\frac{\omega_{pe}^2}{4\pi e}\nabla^2\left(\widetilde{A}a_{DBL}\right),
\end{equation}
where $\widetilde{A}$ and $\tilde{n}_e$ are respectively the vector potential of backscattering light and plasma-density perturbations \cite{kruer1988physics}. For simplicity, we consider a cold plasma, so that the Bohm-Gross frequency for the electron plasma wave $\omega_L=(\omega_{pe}^2+3k_L^2v_{th}^2)^{1/2}\approx\omega_{pe}$, where $v_{th}$ is the thermal velocity. The characteristic time $t_c$ for SRS development is defined as the reciprocal of growth rate when the instability has developed to a considerable level. Without loss of generality, for the strong coupling of the two beamlets, the perturbation of resonance system $\cos(\delta\omega t)\approx1-(\delta\omega t)^2/2$ can be treated as a quasi-static process when $\delta\omega\ll\sqrt{2}/t_c$. Therefore, the dispersion relation of SRS for the two coupled beamlets in the one-dimension (1D) geometry is then obtained as
\begin{equation} \label{4}
\begin{split}
&\frac{\omega^2-\omega_{pe}^2}{\omega_{pe}^2k^2c^2}=\sum_{i=1}^2\frac{a_i^2}{4}\left[\frac{1}{D_{+,i}(k,\omega)}+\frac{1}{D_{-,i}(k,\omega)}\right]\\
&+\frac{a_1a_2}{4}\sum_{i=1}^2\left[\frac{1}{D_{+,i}(k,\omega)}+\frac{1}{D_{-,i}(k,\omega)}\right],
\end{split}
\end{equation}
where $D_{\pm,i}(k,\omega)=(\omega\pm\omega_i)^2-(k\pm k_i)^2c^2-\omega_{pe}^2$. If there is $\delta\omega\gtrsim\sqrt{2}/t_c$, we have the dispersion relation in the decoupling regime as
\begin{equation} \label{5}
\frac{\omega^2-\omega_{pe}^2}{\omega_{pe}^2k^2c^2}=\frac{1}{4}\sum_{i=1}^2\left[\frac{a_i^2}{D_{+,i}(k,\omega)}+\frac{a_i^2}{D_{-,i}(k,\omega)}\right].
\end{equation}
Note that Eqs. (4) and (5) are good approximations in the time scale $t\sim t_c$. An explicit threshold condition for Eq. (5) will be given later.

\begin{figure*}
    \begin{tabular}{lc}
        \begin{overpic}[width=0.7\textwidth]{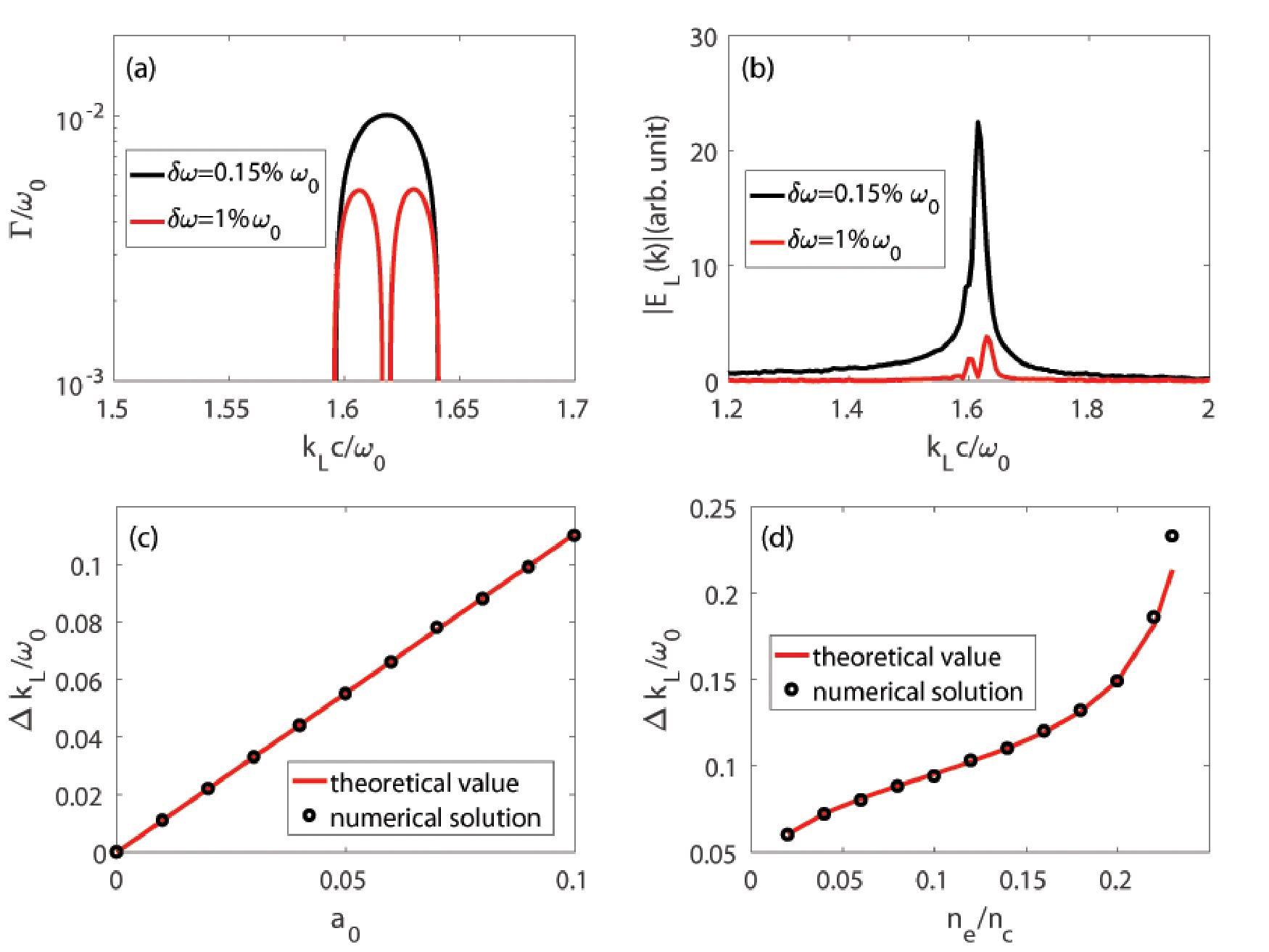}
        \end{overpic}
    \end{tabular}
\caption{ (a) The growth rate $\Gamma$ with $k_L$ for $n_e=0.08n_c$, and $a_1=a_2=0.02$ under two different frequency gaps $\delta\omega$. (b) The corresponding distributions of Langmuir wave vectors at $t=340\tau$ from PIC simulations, where the laser amplitude and plasma density are the same to (a). (c) and (d) show the instability width $\Delta k_L$ of a single laser beam as a function of the laser amplitude and the plasma density, respectively, where the plasma density is $n_e=0.08n_c$ for (c) and the laser amplitude is $a_0=0.08$ for (d). The solid lines are theoretical curves from Eq. (6) and black dots are numerical solutions with Eq. (5).
    }
\end{figure*}

The growth rate of SRS is found by solving Eq. (4) or (5) with the imaginary part of $\omega$, i.e., $\Gamma=\mathrm{Im}(\omega)$, and the area where $\Gamma>0$ is the instability region \cite{zhao2014effects}. Here, taking an example, let us consider the case for the laser amplitudes $a_1=a_2=0.02$ with frequency difference $\delta\omega=0.15\%\omega_0$ and $\delta\omega=1\%\omega_0$. We take the plasma density $n_e=0.08n_c$, where $n_c$ is the critical density. The numerical solutions of the dispersion relation Eqs. (4) and (5) $(\Gamma, k_L)$ with different $\delta\omega$ are plotted in Fig. 2(a). When $\delta\omega=0.15\%\omega_0$, it is found that these the SRS instability regions in the $k_L$ space for the two laser beams overlap to form a single instability region. This implies that two laser pulses are coupled in developing the SRS instability. The wavenumber of the maximum growth rate is $k_L=1.618\omega_0/c$. However, when the frequency difference between the two lasers are increased to $\delta\omega=1\%\omega_0$, the instability regions are separated, each of them will develop independently. Note that the maximum growth rate of the coupled case is much higher than the decoupled one.

To validate the coupling of two lasers, we have carried out PIC simulations by use of {\sc klap} code \cite{chen2008development}. We have taken a homogeneous plasma slab in one-dimension. The length of the simulation box is $200\lambda_0$ where $\lambda_0=2\pi/k_0$, and the plasma occupies a region from $50\lambda_0$ to $150\lambda_0$ with plasma density $n_e=0.08n_c$. The initial temperature is $T_{e0}$ = 100eV. The ions are stationary with a charge $Z=1$. We have taken 100 cells per wavelength and 50 particles per cell. The wavenumber distributions of Langmuir wave are plotted in Fig. 2(b) for $a_1=a_2=0.02$ with two different frequency gaps. Only one peak can be found at $k_L=1.615\omega_0/c$ when the frequency difference $\delta\omega=0.15\%\omega_0$. When $\delta\omega$ increases to $1\%\omega_0$, the strength of Langmuir wave is greatly reduced and two independent peaks can be found at $k_{L1}=1.6\omega_0/c$ and $k_{L2}=1.63\omega_0/c$. This is quite similar as Fig. 2(a). Note that since $\Gamma_{i(max)}\propto k_{Li}$, the strength of the mode with higher $k_L$ is slightly larger. As a result, one can conclude that when the difference of the two laser beams is small enough, they can be coupled with the same plasma wave with a much higher instability growth rate than that corresponding to two individual laser beams.

In the following, we derive a general condition for the decoupling between two lasers. Defining $\Delta k_{L}$ as the width of the instability region for the light with $(k_0,\omega_0)$ and amplitude $a_0$. By letting the growth rate $\Gamma\approx0$ according to Eq. (5) in underdense plasma $n_e<0.25n_c$, one finds
\begin{equation} \label{6}
\Delta k_{L}=a_0
k_{L}\sqrt{\frac{\omega_{pe}(\omega_0-\omega_{pe})}{\omega_0^2-2\omega_0\omega_{pe}}},
\end{equation}
where $k_{L}=k_0+c^{-1}\sqrt{\omega_0^2-2\omega_0\omega_{pe}}$. Solutions of $\Delta k_{L}$ obtained from Eqs. (5) and (6) are compared as shown in Figs. 2(c) and 2(d). One can find that $\Delta k_{L}$ is strictly proportional to the laser amplitude $a_0$ from Fig. 2(c). Based on Fig. 2(d) we know that $\Delta k_{L}$ is also proportional to the plasma density. Generally Eq. (6) fits well with the numerical results of Eq. (5) in the low density regime. In the derivation of Eq. (6), we have assumed that $\omega_{pe}\ll\omega_0$, therefore the theoretical value is smaller than the numerical solution at $n_e>0.23n_c$. The above results indicate that for a given density profile, we can reduce the laser amplitude to shrink the instability region of backward SRS.

In the case of two incident lasers with $(k_0\pm\delta k,\omega_0\pm\delta\omega)$, the plasma wavenumber $k_{L}$ changes with frequency $\omega_0$ according to $dk_{L}/d\omega_0=c^{-1}\omega_0(\omega_0^2-\omega_{pe}^2)^{-1/2}+c^{-1}(\omega_0-\omega_{pe})(\omega_0^2-2\omega_0\omega_{pe})^{-1/2}$. Therefore the condition for decoupling between the two laser beamlets given above in cold plasma can be obtained as $|k_{L1}-k_{L2}|=\delta\omega(d k_L/d\omega_0)>\sqrt{2}\Delta k_L$, where these two instability regions have no intersections in the wavenumber space. When $\omega_{pe}\ll\omega_0$, this simply corresponds to
\begin{equation} \label{7}
\delta\omega/\omega_0>a_0\sqrt{2\omega_{pe}/\omega_0}\approx2\sqrt{2}\Gamma_{SRS}/\omega_0,
\end{equation}
where $\Gamma_{SRS}=(a_0/2)\sqrt{\omega_0\omega_{pe}}$ is the linear growth rate of SRS for a single beamlet with zero bandwidth. Equation (7) defines the required frequency difference for the decoupling of two laser beamlets under the same amplitude $a_0$. In this case, the growth rate is determined by a single beamlet even if the whole laser beam is composed of many beamlets. In this way, the instability of the whole laser beam will be controlled provided the instability of a single beamlet is controlled. This is relatively easy to realize since the energy of a single beamlet can be limited to a low level by increasing the number of beamlets.

Based upon this, we can extend the two laser beamlets to multiple beamlets and define more accurately that a DBL is a light beam composed of many beamlets, where the frequency difference of the neighboring beamlets satisfies Eq. (7). Otherwise if Eq. (7) is not satisfied, we call them as coupled broadband lasers (CBLs). According to Eq. (7), the total bandwidth of the DBL becomes $\Delta\omega>(N-1)\delta\omega$ and the average amplitude of the DBL is $a_{sum}=\sqrt{\sum_{i=1}^N|a_i|^2\omega_i^2/\omega_0^2}$, where $\omega_0$ is the central frequency of the beam.

We point out that some early theoretical study considered the effect of finite laser bandwidth on the instability growth \cite{thomson1974effects,kruer1988physics}. It was proposed that the instability growth rate $\Gamma$ is modified by a laser with finite bandwidth $\delta\omega$ by $\Gamma=\Gamma_{SRS}^2/\delta\omega$. In this case, the linear growth rate is reduced provided $\delta\omega\gg\Gamma_{SRS}=1/t_c$, which can be understood as a destruction of the resonant system. However, this does not imply an effective suppression of the instability when the driving laser energy is high enough.

The frequency difference for decoupling stimulated Brillouin scattering (SBS) can be obtained in a similar way as the above SRS case. Under $a_0\lesssim10^{-2}$ and $n_e\ll n_c$, it can be reduced to
\begin{equation} \label{8}
\delta\omega/\omega_0>2a_0\frac{\omega_{pi}}{\omega_0}\sqrt{\frac{\omega_0}{k_0c_s}}\approx4\sqrt{2}\Gamma_{SBS}/\omega_0,
\end{equation}
where $\omega_{pi}=\omega_{pe}\sqrt{Zm_e/m_i}$, and $\Gamma_{SBS}$ is the linear growth rate of SBS for a single beamlet with zero bandwidth. Under indirect-drive conditions $c_s\sim10^{-2}c$ and $n_e\lesssim0.1n_c$, the decoupled threshold for $\delta \omega$ given in Eq. (7) is always larger than Eq. (8), therefore once Eq. (7) is satisfied, both SRS and SBS can be suppressed.

\section{PIC simulations of interactions between DBLs and plasmas}
\subsection{In homogeneous plasma}

Following above discussion, we will demonstrate explicitly the effects of DBLs on the suppression of SRS instabilities by use of some examples. The laser beams are given in the form of Eq. (1), where the amplitude of each laser beamlet $a_i$ is set to be a constant. Firstly, we compare two cases of light beams, one is a CBL and the other is DBL, each composed of $N=100$ beamlets. The amplitudes for the two cases are respectively $a_i=0.004$ and $a_i=0.001$ for i=1 to 100. The total light energy of the 100 beams is the same as the single beam with the amplitude $a_{sum}=0.04$ and $a_{sum}=0.01$, respectively. The frequency difference between neighboring beamlets is fixed as $\delta\omega=0.12\%\omega_0$. Then the overall bandwidth can be obtained according to $\Delta\omega_0=(N-1)\delta\omega\approx12\%\omega_0$ for both beams. We take the homogeneous plasma density $n_0=0.08n_c$. According to Eq. (7), the above plasma density and laser bandwidth suggest that the amplitude for each beamlet must be less than $a_i=0.0016$ in order to suppression the coupling between neighboring beamlets and overall development of SRS. Therefore, the case with $a_i=0.004$ corresponds to a CBL, where SRS cannot be suppressed, and the case with $a_i=0.001$ is a DBL, where SRS can be effectively suppressed.

\begin{figure}
    \begin{tabular}{lc}
        \begin{overpic}[width=0.35\textwidth]{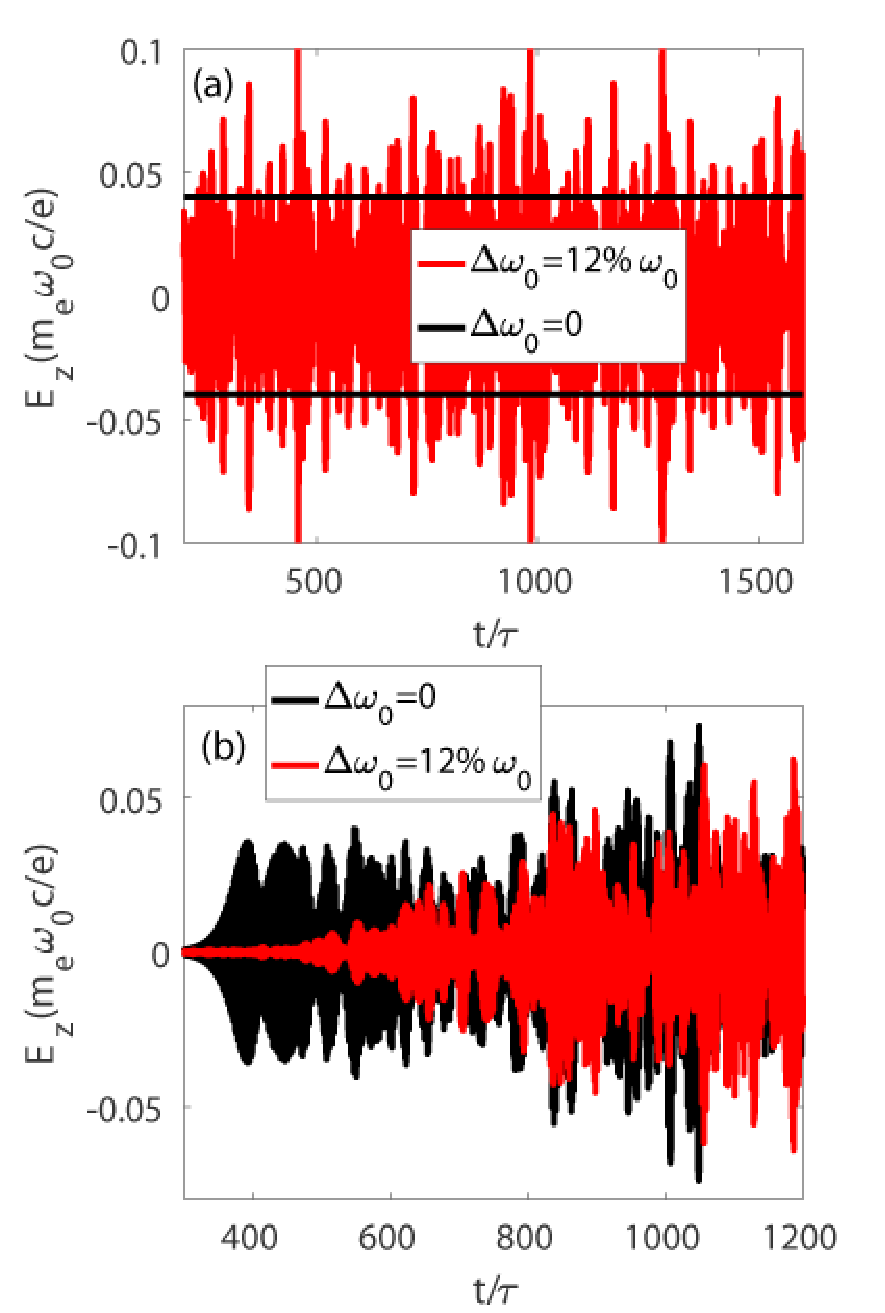}
        \end{overpic}
    \end{tabular}
\caption{ (a) Temporal envelopes of the incident lights for a normal laser beam and a CBL at the amplitude $a_{sum}=0.04$, where the total bandwidth of the CBL is 12\%. The black solid lines indicate the laser field amplitude level for the normal laser beam. (b) Temporal evolution of backscattering light developed by normal laser and CBL.
    }
\end{figure}

To validate the above theory prediction, series of 1D PIC simulations have been performed in homogeneous plasma for the interactions between laser beams and plasma. The plasma length is $400\lambda$ and some vacuum regions are set at the two side of the plasma region. Figure 3(a) shows an example of the temporal structure when taking $a_i=0.004$, $\delta\omega=0.12\%\omega_0$, and $N=100$, which is the CBL mentioned above. It shows that there are some fluctuations in the envelope profile. But overall the amplitude appears around $a_{sum}=(\sum_{i=1}^N|a_{i}|^2|\omega_i|^2)^{1/2}=0.04$. It is to be compared to the single coherent laser beam with the same amplitude $a_{sum}=0.04$, $\Delta\omega_0=0$. The temporal envelop of the backscattered light is shown in Fig. 3(b). The growth rate of backscattered light with the CBL is considerably reduced as compared to a normal laser beam. However, after certain time about $t=500\tau$, the backscattered light starts to grow quickly. Theoretically the growth rate for a beam with $a_i=0.004$ is found to be $\Gamma_{max}=0.0016\omega_0$, and its characteristic time is $t_c=1/\Gamma_{max}=625\tau$ with $\tau$ the laser oscillation period. The coupling between neighboring beamlets leads to a higher growth rate and a high SRS level. Finally at $t=800\tau$, the scattered light saturates at the same level as produced by normal coherent lasers. From these results, one concludes that instability can grow to a high level due to the coupling between neighboring beamlets for a CBL even if the overall bandwidth is very high.

\begin{figure*}
    \begin{tabular}{lc}
        \begin{overpic}[width=0.7\textwidth]{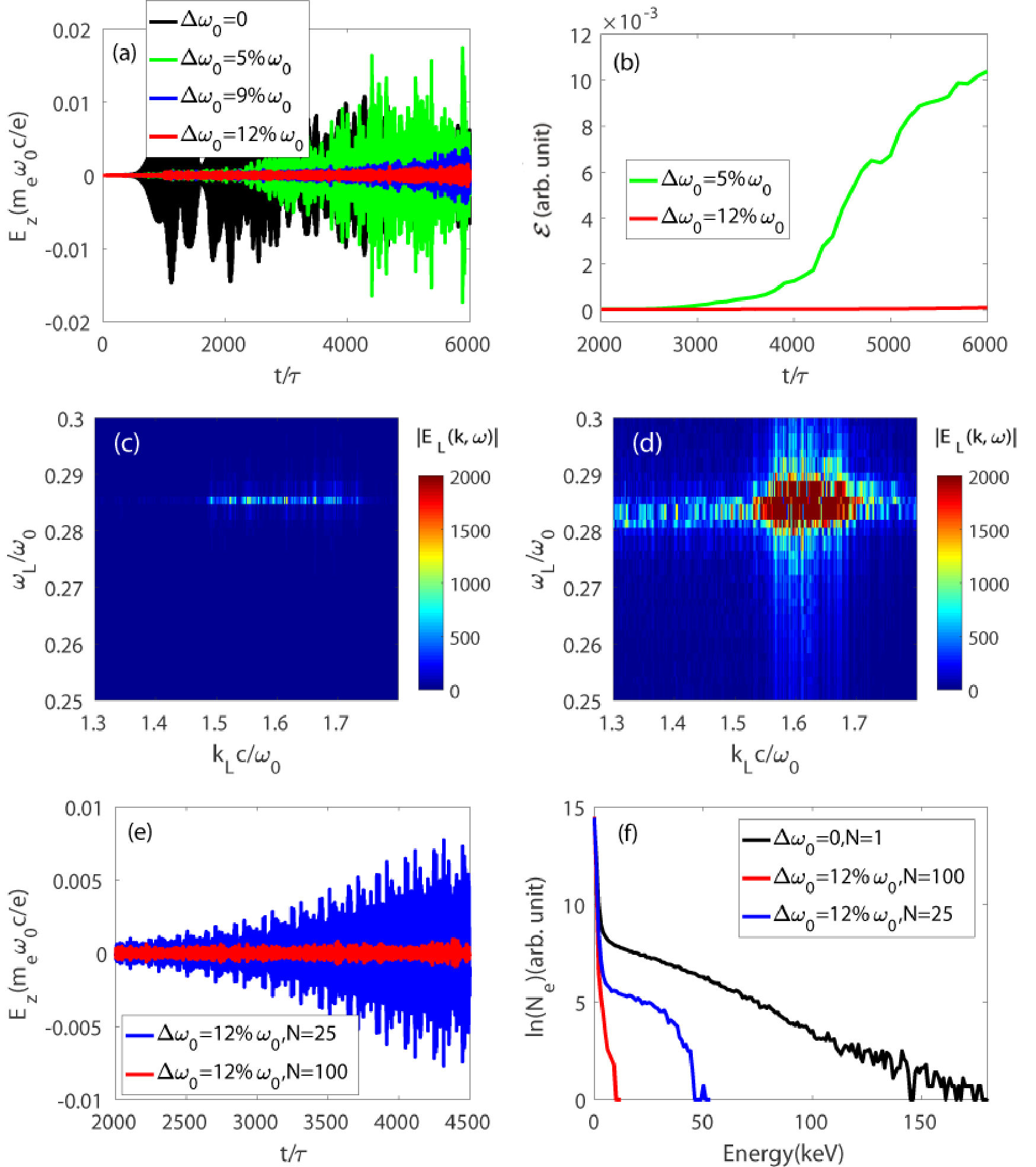}
        \end{overpic}
    \end{tabular}
\caption{ (a) Temporal profiles of the backscattering light found for the incident light with different bandwidths under the same energy. (b) Temporal evolution of electrostatic energy for the incident light with different bandwidth. (c) and (d) Distributions of the Langmuir wave in $(k_L,\omega_L)$ space obtained for the time window $[5500,6000]\tau$ with bandwidth $\Delta\omega_0=12\%\omega_0$ and $\Delta\omega_0=5\%\omega_0$, respectively. (e) Temporal profiles of the backscattering light found respectively for the DBL composed of 25 or 100 beamlets under the same light energy and bandwidth $\Delta\omega=12\%\omega_0$. (f) Energy distributions of electrons found respectively for the normal laser beam, and the DBL with $\Delta\omega=12\%\omega_0$ and different beam number $N$ under the same light energy. $N_e$ is the relative electron number.
    }
\end{figure*}

Now we consider the opposite case with a DBL for $a_i=0.001$ and the same beam number $N=100$. At this amplitude, the decoupling threshold for the bandwidth is about $\delta\omega >0.08\%\omega_0$ or $\Delta\omega>8\%\omega_0$ according to Eq. (7). When the threshold is satisfied, the growth of backscattering light is greatly reduced, as shown in Fig. 4(a), where a comparison between different bandwidth cases is made. The backscattered SRS light is found at a very low level when the bandwidth is larger than the threshold. The maximum amplitude for the DBL with $\Delta\omega_0=12\%\omega_0$ at $t=6000\tau$ is $E_z=0.0016$, which is about an order of magnitude smaller than the case for the normal laser beam. Therefore, the electron heating is almost completely suppressed at $t=6000\tau$ as shown in Fig. 4(f). On the contrary, for the case with a normal laser light, the backscattering light reaches to a high saturation level quickly. Correspondingly, hot electrons with temperature around $T_e=16.6$keV are generated, which corresponds to electron heating by the large amplitude of Langmuir wave with a phase velocity about $v_{ph}=0.18c$. We diagnose the energy of the Langmuir wave $\mathcal{E}=\int E_L^2dx$ which is a direct estimation of the strength of SRS. One finds that $\mathcal{E}$ does not show to grow at all for the DBL with $\Delta\omega_0=12\%\omega_0$. On the contrary, $\mathcal{E}$ increases exponentially at $t=3000\tau$ when $\Delta\omega_0=5\%\omega_0$ as shown in Fig. 4(b).

Figures 4(c) and 4(d) present the Langmuir wave in the $(k_L,\omega_L)$ space with different bandwidth. Considering the incident light frequency changes in [0.94,1.06]$\omega_0$, one finds that the corresponding $k_L$ ranges in [1.49,1.74]$\omega_0/c$, taking $\omega_{pe}=0.283\omega_0$. The intensity of comb-spectrum for $\Delta\omega_0=12\%\omega_0$ is much weaker than the $\Delta\omega_0=5\%\omega_0$ case, as shown in the comparison between Fig. 4(c) and 4(d). One finds a strong coupling of beamlets around $\omega_{pe}=0.283\omega_0$ for the latter one.

One notes that the overall SRS development of a DBL depends upon two factors: the coupling/decoupling between neighboring beamlets and the development of SRS with a single beamlet. These cause different behaviors of SRS development for a DBL composed of different number of beamlets $N$, even though the overall bandwidth and laser intensity are the same. For example, for the same amplitude $a_{sum}=0.01$ and the bandwidth $\Delta\omega_0=12\%\omega_0$, Fig. 4(e) compares the scattering light as a function of time for $N=25$ and 100. For both cases, the decoupling condition given by Eq. (7) is satisfied. However, SRS is found to develop much quickly for the case with $N=25$ since the amplitude and the SRS growth rate of each beamlet are larger than that for $N=100$. The larger growth rate of $N=25$ case leads to the production of hot electrons as shown in Fig. 4(f).

\begin{figure}
    \begin{tabular}{lc}
        \begin{overpic}[width=0.35\textwidth]{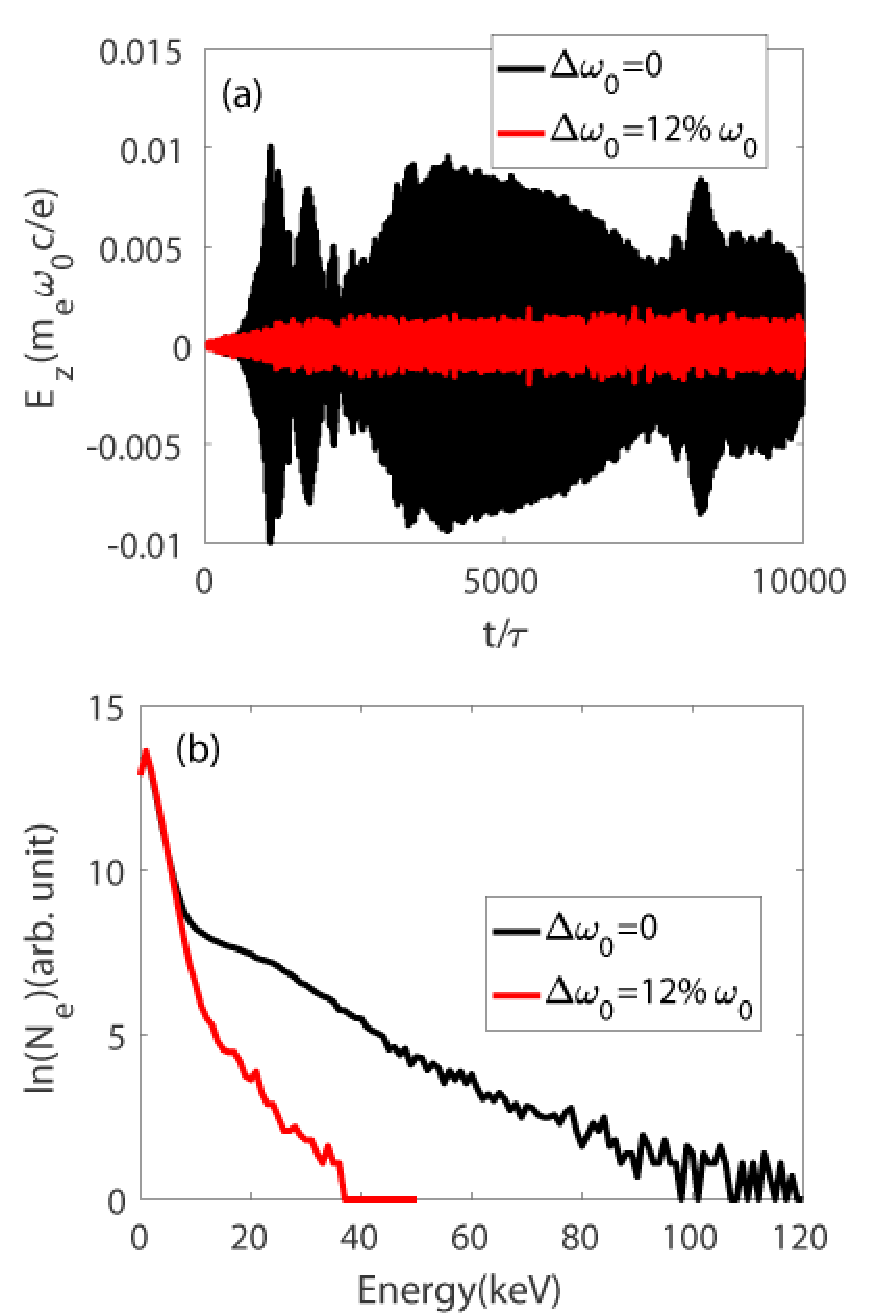}
        \end{overpic}
    \end{tabular}
\caption{ (a) Temporal profiles of the backscattering light found respectively for the normal laser beam with amplitude $a_{sum}=0.01$ and the DBL composed of 100 beams each with $a_i=0.001$ under different bandwidths. (b) Energy distributions of electrons found respectively for the normal laser beam and the DBL with $\Delta\omega=12\%\omega_0$, at $t=6000\tau$. The initial electron temperature is $T_{e0}$ = 1keV with mobile ions.
    }
\end{figure}

In a hot plasma, Landau damping provides a threshold for the onset of instabilities, therefore the suppression of SRS is more effective. Here a simulation for 1keV hot plasma with mobile ions is performed up to 10000$\tau$. From Fig. 5(a) we know that a large amplitude of backscattered light is produced by normal laser through the development of SRS and SBS. The strong SRS leads to large numbers of hot electrons as shown in Fig. 5(b). By contrast to it, both SRS and SBS have not been obviously developed during 10000$\tau$ in the case with DBLs. As we discussed in Sec. II, the DBL can also suppress SBS when the frequency difference satisfies Eq. (7). The above simulations imply that DBLs can overcome the two major problems (laser energy loss and hot electron production) in laser plasma interactions.

\subsection{Effect of nonuniform plasma density}

The above theory and simulation are developed for homogeneous plasma. It is expected that the SRS suppression with DBLs is also effective in inhomogeneous plasma. Assuming an inhomogeneous plasma density profile $n_e(x)=n_0(1+x/L_n)$, where $L_n\sim$mm inside a Hohlraum target for indirect-drive ICF \cite{myatt2014multiple}. For an inhomogeneous plasma, the coupling of each beams will be reduced when their resonant region $\Delta x=4\Gamma/(K'\sqrt{v_1v_2})$ decreased, where $K'\propto\omega_{pe}/L_n$, $v_1$ and $v_2$ are the group velocity of scattered light and Langmuir wave, respectively \cite{liu1974raman}. When $L_n$ approaches to infinite, the situation transits to homogeneous case. Therefore, the convective instability can be more easily suppressed when $a_i$ or $L_n$ is reduced. An upper-limit threshold is provided by Eq. (7) for inhomogeneous plasma. For NIF with the peak laser intensity $I=8\times10^{14}\mathrm{W/cm^2}$ and laser wavelength $\lambda=0.35\mu$m, the corresponding laser amplitude is $a_0\sim0.0085$. 1D PIC simulations were performed in inhomogeneous plasma with $n_0=0.07n_c$, $L_n=3000\lambda_0$, and the plasma density linearly ranges in [0.07,0.09]$n_c$. The initial electron temperature is $T_{e0}=100$eV. To compare with the above simulation, here we take $a_i=0.001$ and $N=100$.

\begin{figure}
\centering
    \begin{tabular}{lc}
        \begin{overpic}[width=0.35\textwidth]{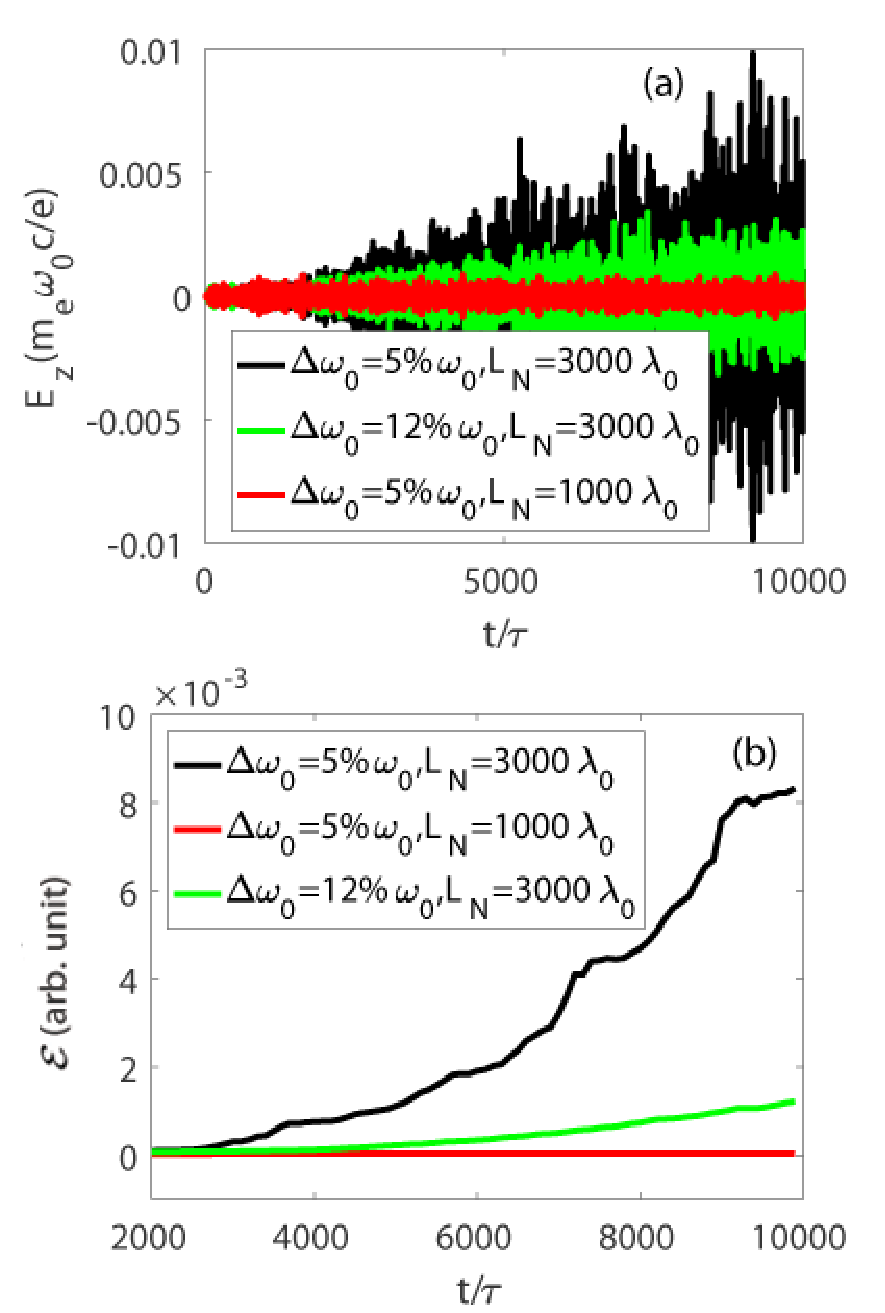}
        \end{overpic}
    \end{tabular}
\caption{ Temporal evolution of (a) backscattering light and (b) electrostatic energy for $a_i=0.001$ with different bandwidth or density gradient $L_n$. The initial electron temperature is $T_{e0}$ = 100eV.
    }
\end{figure}

The envelop of backscattering light with different bandwidth or density gradient is presented in Fig. 6(a). When $L_n=3000\lambda_0$, the condition is close to the NIF situations. For the light with $\Delta\omega_0=12\%\omega_0$, $E_z$ grows with a very small growth rate. On the contrary, a large amplitude of backscattering light is produced when $\Delta\omega_0=5\%\omega_0$. These results are similar to the homogeneous case. If $L_n$ decreased to $1000\lambda_0$, the coupling of DBLs is reduced due to the resonant region becomes narrow, which leads to the complete suppression of SRS.

Figure 6(b) shows the evolutions of $\mathcal{E}$ for coherent laser and DBL, the results are similar to Fig. 6(a). At $L_n=3000\lambda_0$, $\mathcal{E}$ grows linearly with a very small growth rate when $\Delta\omega_0=12\%\omega_0$. On the contrary, $\mathcal{E}$ increases exponentially at $t=3000\tau$ when $\Delta\omega_0=5\%\omega_0$, and large numbers of hot electrons are produced at $t=7000\tau$. If $L_n$ decreased to $1000\lambda_0$, SRS is almost completely suppressed due to the decoupling of beamlets and the large threshold for the onset of instabilities as shown in Fig. 6(b).

In passing, we mention that, even though we have only shown the effectiveness of SRS suppression with DBLs with 1D simulation, it is also true in multi-dimensional cases. This is because typically the backscattering has the highest growth rate than the side scattering. Once the backscattering is suppressed, side scattering will be controlled as demonstrated by our 2D simulation \cite{zhao2015effects}.

\section{Summary and discussion}

In conclusion, we have proposed a strategy to suppress SRS significantly by use of so called DBLs with certain bandwidth. It is based upon a model of the coupling between two laser beams with slightly different frequencies. It is found that the couple of the two laser beams in the excitation of SRS is weak as long as their frequency difference is larger than the 70\% width of instability region for an individual beam. The latter is proportional to the laser amplitude. Therefore, with a DBL composed of many beamlets (such as 100) with certain frequency difference between individual beamlets (such as 0.12\%$\omega_0$), SRS can be dramatically suppressed due to the decoupling of the beamlets. Since the required bandwidth of a DBL for SBS suppression is typically smaller than that for SRS, SBS will be suppressed simultaneously as long as SRS is suppressed.

It is expected that the DBLs may also be applied to suppress other parametric instabilities for ICF applications, such as the suppression of two-plasmon decay (TPD) instabilities. Near the quarter critical density, absolute SRS has the largest growth rate, it is thus expected that the required laser bandwidth has to be increased under the same laser intensity. Note that the TPD instability growth is comparable to the absolute SRS, therefore a DBL with larger bandwidth is needed to suppress TPD. Generally, the laser technology for DBLs still needs to be developed. Note that the comb-like spectrum for DBLs can be produced with different schemes \cite{Ho1993,Yu2016}. Also the gain bandwidth of lasers over 10\% can be realised via parametric amplification in nonlinear crystals \cite{Dabu2010}. Therefore, in principle it is possible to build a high power laser system for DBLs.

\section{Acknowledgement}

This work was supported by National Science Foundation of China (Grant No. 11421064, 11129503 and 11775144).

\bibliographystyle{apsrev4-1}

\begin{thebibliography}{35}%
\makeatletter
\providecommand \@ifxundefined [1]{%
 \@ifx{#1\undefined}
}%
\providecommand \@ifnum [1]{%
 \ifnum #1\expandafter \@firstoftwo
 \else \expandafter \@secondoftwo
 \fi
}%
\providecommand \@ifx [1]{%
 \ifx #1\expandafter \@firstoftwo
 \else \expandafter \@secondoftwo
 \fi
}%
\providecommand \natexlab [1]{#1}%
\providecommand \enquote  [1]{``#1''}%
\providecommand \bibnamefont  [1]{#1}%
\providecommand \bibfnamefont [1]{#1}%
\providecommand \citenamefont [1]{#1}%
\providecommand \href@noop [0]{\@secondoftwo}%
\providecommand \href [0]{\begingroup \@sanitize@url \@href}%
\providecommand \@href[1]{\@@startlink{#1}\@@href}%
\providecommand \@@href[1]{\endgroup#1\@@endlink}%
\providecommand \@sanitize@url [0]{\catcode `\\12\catcode `\$12\catcode
  `\&12\catcode `\#12\catcode `\^12\catcode `\_12\catcode `\%12\relax}%
\providecommand \@@startlink[1]{}%
\providecommand \@@endlink[0]{}%
\providecommand \url  [0]{\begingroup\@sanitize@url \@url }%
\providecommand \@url [1]{\endgroup\@href {#1}{\urlprefix }}%
\providecommand \urlprefix  [0]{URL }%
\providecommand \Eprint [0]{\href }%
\providecommand \doibase [0]{http://dx.doi.org/}%
\providecommand \selectlanguage [0]{\@gobble}%
\providecommand \bibinfo  [0]{\@secondoftwo}%
\providecommand \bibfield  [0]{\@secondoftwo}%
\providecommand \translation [1]{[#1]}%
\providecommand \BibitemOpen [0]{}%
\providecommand \bibitemStop [0]{}%
\providecommand \bibitemNoStop [0]{.\EOS\space}%
\providecommand \EOS [0]{\spacefactor3000\relax}%
\providecommand \BibitemShut  [1]{\csname bibitem#1\endcsname}%
\let\auto@bib@innerbib\@empty
\bibitem [{\citenamefont {Betti}\ and\ \citenamefont
  {Hurricane}(2016)}]{Betti2016Inertial}%
  \BibitemOpen
  \bibfield  {author} {\bibinfo {author} {\bibfnamefont {R.}~\bibnamefont
  {Betti}}\ and\ \bibinfo {author} {\bibfnamefont {O.~A.}\ \bibnamefont
  {Hurricane}},\ }\href@noop {} {\bibfield  {journal} {\bibinfo  {journal}
  {Nature Phys.}\ }\textbf {\bibinfo {volume} {12}},\ \bibinfo {pages} {435}
  (\bibinfo {year} {2016})}\BibitemShut {NoStop}%
\bibitem [{\citenamefont {Lindl}\ \emph {et~al.}(2014)\citenamefont {Lindl},
  \citenamefont {Landen}, \citenamefont {Edwards}, \citenamefont {Moses},\ and\
  \citenamefont {Team}}]{Lindl2014Review}%
  \BibitemOpen
  \bibfield  {author} {\bibinfo {author} {\bibfnamefont {J.}~\bibnamefont
  {Lindl}}, \bibinfo {author} {\bibfnamefont {O.}~\bibnamefont {Landen}},
  \bibinfo {author} {\bibfnamefont {J.}~\bibnamefont {Edwards}}, \bibinfo
  {author} {\bibfnamefont {E.}~\bibnamefont {Moses}}, \ and\ \bibinfo {author}
  {\bibfnamefont {N.}~\bibnamefont {Team}},\ }\href@noop {} {\bibfield
  {journal} {\bibinfo  {journal} {Phys. Plasmas}\ }\textbf {\bibinfo {volume}
  {21}},\ \bibinfo {pages} {020501} (\bibinfo {year} {2014})}\BibitemShut
  {NoStop}%
\bibitem [{\citenamefont {Hurricane}\ \emph {et~al.}(2014)\citenamefont
  {Hurricane}, \citenamefont {Callahan}, \citenamefont {Casey}, \citenamefont
  {Celliers}, \citenamefont {Cerjan}, \citenamefont {Dewald}, \citenamefont
  {Dittrich}, \citenamefont {D{\"o}ppner}, \citenamefont {Hinkel},\ and\
  \citenamefont {Berzak~Hopkins}}]{Hurricane2014Fuel}%
  \BibitemOpen
  \bibfield  {author} {\bibinfo {author} {\bibfnamefont {O.~A.}\ \bibnamefont
  {Hurricane}}, \bibinfo {author} {\bibfnamefont {D.~A.}\ \bibnamefont
  {Callahan}}, \bibinfo {author} {\bibfnamefont {D.~T.}\ \bibnamefont {Casey}},
  \bibinfo {author} {\bibfnamefont {P.~M.}\ \bibnamefont {Celliers}}, \bibinfo
  {author} {\bibfnamefont {C.}~\bibnamefont {Cerjan}}, \bibinfo {author}
  {\bibfnamefont {E.~L.}\ \bibnamefont {Dewald}}, \bibinfo {author}
  {\bibfnamefont {T.~R.}\ \bibnamefont {Dittrich}}, \bibinfo {author}
  {\bibfnamefont {T.}~\bibnamefont {D{\"o}ppner}}, \bibinfo {author}
  {\bibfnamefont {D.~E.}\ \bibnamefont {Hinkel}}, \ and\ \bibinfo {author}
  {\bibfnamefont {L.~F.}\ \bibnamefont {Berzak~Hopkins}},\ }\href@noop {}
  {\bibfield  {journal} {\bibinfo  {journal} {Nature}\ }\textbf {\bibinfo
  {volume} {506}},\ \bibinfo {pages} {343} (\bibinfo {year}
  {2014})}\BibitemShut {NoStop}%
\bibitem [{\citenamefont {Town}\ \emph {et~al.}(2014)\citenamefont {Town},
  \citenamefont {Bradley}, \citenamefont {Kritcher}, \citenamefont {Jones},
  \citenamefont {Rygg}, \citenamefont {Tommasini}, \citenamefont {Barrios},
  \citenamefont {Benedetti}, \citenamefont {Hopkins},\ and\ \citenamefont
  {Celliers}}]{Town2014Dynamic}%
  \BibitemOpen
  \bibfield  {author} {\bibinfo {author} {\bibfnamefont {R.~P.~J.}\
  \bibnamefont {Town}}, \bibinfo {author} {\bibfnamefont {D.~K.}\ \bibnamefont
  {Bradley}}, \bibinfo {author} {\bibfnamefont {A.}~\bibnamefont {Kritcher}},
  \bibinfo {author} {\bibfnamefont {O.~S.}\ \bibnamefont {Jones}}, \bibinfo
  {author} {\bibfnamefont {J.~R.}\ \bibnamefont {Rygg}}, \bibinfo {author}
  {\bibfnamefont {R.}~\bibnamefont {Tommasini}}, \bibinfo {author}
  {\bibfnamefont {M.}~\bibnamefont {Barrios}}, \bibinfo {author} {\bibfnamefont
  {L.~R.}\ \bibnamefont {Benedetti}}, \bibinfo {author} {\bibfnamefont
  {L.~F.~B.}\ \bibnamefont {Hopkins}}, \ and\ \bibinfo {author} {\bibfnamefont
  {P.~M.}\ \bibnamefont {Celliers}},\ }\href@noop {} {\bibfield  {journal}
  {\bibinfo  {journal} {Phys. Plasmas}\ }\textbf {\bibinfo {volume} {21}},\
  \bibinfo {pages} {041006} (\bibinfo {year} {2014})}\BibitemShut {NoStop}%
\bibitem [{\citenamefont {Moody}\ \emph
  {et~al.}(2014{\natexlab{a}})\citenamefont {Moody}, \citenamefont {Robey},
  \citenamefont {Celliers}, \citenamefont {Munro}, \citenamefont {Barker},
  \citenamefont {Baker}, \citenamefont {D{\"o}ppner}, \citenamefont {Hash},
  \citenamefont {Hopkins},\ and\ \citenamefont {Lafortune}}]{Moody2014Early}%
  \BibitemOpen
  \bibfield  {author} {\bibinfo {author} {\bibfnamefont {J.~D.}\ \bibnamefont
  {Moody}}, \bibinfo {author} {\bibfnamefont {H.~F.}\ \bibnamefont {Robey}},
  \bibinfo {author} {\bibfnamefont {P.~M.}\ \bibnamefont {Celliers}}, \bibinfo
  {author} {\bibfnamefont {D.~H.}\ \bibnamefont {Munro}}, \bibinfo {author}
  {\bibfnamefont {D.~A.}\ \bibnamefont {Barker}}, \bibinfo {author}
  {\bibfnamefont {K.~L.}\ \bibnamefont {Baker}}, \bibinfo {author}
  {\bibfnamefont {T.}~\bibnamefont {D{\"o}ppner}}, \bibinfo {author}
  {\bibfnamefont {N.~L.}\ \bibnamefont {Hash}}, \bibinfo {author}
  {\bibfnamefont {L.~B.}\ \bibnamefont {Hopkins}}, \ and\ \bibinfo {author}
  {\bibfnamefont {K.}~\bibnamefont {Lafortune}},\ }\href@noop {} {\bibfield
  {journal} {\bibinfo  {journal} {Phys. Plasmas}\ }\textbf {\bibinfo {volume}
  {21}},\ \bibinfo {pages} {092702} (\bibinfo {year}
  {2014}{\natexlab{a}})}\BibitemShut {NoStop}%
\bibitem [{\citenamefont {Moody}\ \emph
  {et~al.}(2014{\natexlab{b}})\citenamefont {Moody}, \citenamefont {Callahan},
  \citenamefont {Hinkel}, \citenamefont {Amendt}, \citenamefont {Baker},
  \citenamefont {Bradley}, \citenamefont {Celliers}, \citenamefont {Dewald},
  \citenamefont {Divol},\ and\ \citenamefont
  {D{\"o}ppner}}]{Moody2014Progress}%
  \BibitemOpen
  \bibfield  {author} {\bibinfo {author} {\bibfnamefont {J.~D.}\ \bibnamefont
  {Moody}}, \bibinfo {author} {\bibfnamefont {D.~A.}\ \bibnamefont {Callahan}},
  \bibinfo {author} {\bibfnamefont {D.~E.}\ \bibnamefont {Hinkel}}, \bibinfo
  {author} {\bibfnamefont {P.~A.}\ \bibnamefont {Amendt}}, \bibinfo {author}
  {\bibfnamefont {K.~L.}\ \bibnamefont {Baker}}, \bibinfo {author}
  {\bibfnamefont {D.}~\bibnamefont {Bradley}}, \bibinfo {author} {\bibfnamefont
  {P.~M.}\ \bibnamefont {Celliers}}, \bibinfo {author} {\bibfnamefont {E.~L.}\
  \bibnamefont {Dewald}}, \bibinfo {author} {\bibfnamefont {L.}~\bibnamefont
  {Divol}}, \ and\ \bibinfo {author} {\bibfnamefont {T.}~\bibnamefont
  {D{\"o}ppner}},\ }\href@noop {} {\bibfield  {journal} {\bibinfo  {journal}
  {Phys. Plasmas}\ }\textbf {\bibinfo {volume} {21}},\ \bibinfo {pages}
  {056317} (\bibinfo {year} {2014}{\natexlab{b}})}\BibitemShut {NoStop}%
\bibitem [{\citenamefont {Igumenshchev}\ \emph {et~al.}(2012)\citenamefont
  {Igumenshchev}, \citenamefont {Seka}, \citenamefont {Edgell}, \citenamefont
  {Michel}, \citenamefont {Froula}, \citenamefont {Goncharov}, \citenamefont
  {Craxton}, \citenamefont {Divol}, \citenamefont {Epstein}, \citenamefont
  {Follett}, \citenamefont {Kelly}, \citenamefont {Kosc}, \citenamefont
  {Maximov}, \citenamefont {McCrory}, \citenamefont {Meyerhofer}, \citenamefont
  {Michel}, \citenamefont {Myatt}, \citenamefont {Sangster}, \citenamefont
  {Shvydky}, \citenamefont {Skupsky},\ and\ \citenamefont
  {Stoeckl}}]{Igumenshchev2012Crossed}%
  \BibitemOpen
  \bibfield  {author} {\bibinfo {author} {\bibfnamefont {I.~V.}\ \bibnamefont
  {Igumenshchev}}, \bibinfo {author} {\bibfnamefont {W.}~\bibnamefont {Seka}},
  \bibinfo {author} {\bibfnamefont {D.~H.}\ \bibnamefont {Edgell}}, \bibinfo
  {author} {\bibfnamefont {D.~T.}\ \bibnamefont {Michel}}, \bibinfo {author}
  {\bibfnamefont {D.~H.}\ \bibnamefont {Froula}}, \bibinfo {author}
  {\bibfnamefont {V.~N.}\ \bibnamefont {Goncharov}}, \bibinfo {author}
  {\bibfnamefont {R.~S.}\ \bibnamefont {Craxton}}, \bibinfo {author}
  {\bibfnamefont {L.}~\bibnamefont {Divol}}, \bibinfo {author} {\bibfnamefont
  {R.}~\bibnamefont {Epstein}}, \bibinfo {author} {\bibfnamefont
  {R.}~\bibnamefont {Follett}}, \bibinfo {author} {\bibfnamefont {J.~H.}\
  \bibnamefont {Kelly}}, \bibinfo {author} {\bibfnamefont {T.~Z.}\ \bibnamefont
  {Kosc}}, \bibinfo {author} {\bibfnamefont {A.~V.}\ \bibnamefont {Maximov}},
  \bibinfo {author} {\bibfnamefont {R.~L.}\ \bibnamefont {McCrory}}, \bibinfo
  {author} {\bibfnamefont {D.~D.}\ \bibnamefont {Meyerhofer}}, \bibinfo
  {author} {\bibfnamefont {P.}~\bibnamefont {Michel}}, \bibinfo {author}
  {\bibfnamefont {J.~F.}\ \bibnamefont {Myatt}}, \bibinfo {author}
  {\bibfnamefont {T.~C.}\ \bibnamefont {Sangster}}, \bibinfo {author}
  {\bibfnamefont {A.}~\bibnamefont {Shvydky}}, \bibinfo {author} {\bibfnamefont
  {S.}~\bibnamefont {Skupsky}}, \ and\ \bibinfo {author} {\bibfnamefont
  {C.}~\bibnamefont {Stoeckl}},\ }\href@noop {} {\bibfield  {journal} {\bibinfo
   {journal} {Phys. Plasmas}\ }\textbf {\bibinfo {volume} {19}},\ \bibinfo
  {pages} {056314} (\bibinfo {year} {2012})}\BibitemShut {NoStop}%
\bibitem [{\citenamefont {Regan}\ \emph {et~al.}(2010)\citenamefont {Regan},
  \citenamefont {Meezan}, \citenamefont {Suter}, \citenamefont {Strozzi},
  \citenamefont {Kruer}, \citenamefont {Meeker}, \citenamefont {Glenzer},
  \citenamefont {Seka}, \citenamefont {Stoeckl},\ and\ \citenamefont
  {Glebov}}]{Regan2010Suprathermal}%
  \BibitemOpen
  \bibfield  {author} {\bibinfo {author} {\bibfnamefont {S.~P.}\ \bibnamefont
  {Regan}}, \bibinfo {author} {\bibfnamefont {N.~B.}\ \bibnamefont {Meezan}},
  \bibinfo {author} {\bibfnamefont {L.~J.}\ \bibnamefont {Suter}}, \bibinfo
  {author} {\bibfnamefont {D.~J.}\ \bibnamefont {Strozzi}}, \bibinfo {author}
  {\bibfnamefont {W.~L.}\ \bibnamefont {Kruer}}, \bibinfo {author}
  {\bibfnamefont {D.}~\bibnamefont {Meeker}}, \bibinfo {author} {\bibfnamefont
  {S.~H.}\ \bibnamefont {Glenzer}}, \bibinfo {author} {\bibfnamefont
  {W.}~\bibnamefont {Seka}}, \bibinfo {author} {\bibfnamefont {C.}~\bibnamefont
  {Stoeckl}}, \ and\ \bibinfo {author} {\bibfnamefont {V.~Y.}\ \bibnamefont
  {Glebov}},\ }\href@noop {} {\bibfield  {journal} {\bibinfo  {journal} {Phys.
  Plasmas}\ }\textbf {\bibinfo {volume} {17}},\ \bibinfo {pages} {055503}
  (\bibinfo {year} {2010})}\BibitemShut {NoStop}%
\bibitem [{\citenamefont {Smalyuk}\ \emph {et~al.}(2008)\citenamefont
  {Smalyuk}, \citenamefont {Shvarts}, \citenamefont {Betti}, \citenamefont
  {Delettrez}, \citenamefont {Edgell}, \citenamefont {Glebov}, \citenamefont
  {Goncharov}, \citenamefont {McCrory}, \citenamefont {Meyerhofer},
  \citenamefont {Radha}, \citenamefont {Regan}, \citenamefont {Sangster},
  \citenamefont {Seka}, \citenamefont {Skupsky}, \citenamefont {Stoeckl},
  \citenamefont {Yaakobi}, \citenamefont {Frenje}, \citenamefont {Li},
  \citenamefont {Petrasso},\ and\ \citenamefont {S\'eguin}}]{Smalyuk2008Role}%
  \BibitemOpen
  \bibfield  {author} {\bibinfo {author} {\bibfnamefont {V.~A.}\ \bibnamefont
  {Smalyuk}}, \bibinfo {author} {\bibfnamefont {D.}~\bibnamefont {Shvarts}},
  \bibinfo {author} {\bibfnamefont {R.}~\bibnamefont {Betti}}, \bibinfo
  {author} {\bibfnamefont {J.~A.}\ \bibnamefont {Delettrez}}, \bibinfo {author}
  {\bibfnamefont {D.~H.}\ \bibnamefont {Edgell}}, \bibinfo {author}
  {\bibfnamefont {V.~Y.}\ \bibnamefont {Glebov}}, \bibinfo {author}
  {\bibfnamefont {V.~N.}\ \bibnamefont {Goncharov}}, \bibinfo {author}
  {\bibfnamefont {R.~L.}\ \bibnamefont {McCrory}}, \bibinfo {author}
  {\bibfnamefont {D.~D.}\ \bibnamefont {Meyerhofer}}, \bibinfo {author}
  {\bibfnamefont {P.~B.}\ \bibnamefont {Radha}}, \bibinfo {author}
  {\bibfnamefont {S.~P.}\ \bibnamefont {Regan}}, \bibinfo {author}
  {\bibfnamefont {T.~C.}\ \bibnamefont {Sangster}}, \bibinfo {author}
  {\bibfnamefont {W.}~\bibnamefont {Seka}}, \bibinfo {author} {\bibfnamefont
  {S.}~\bibnamefont {Skupsky}}, \bibinfo {author} {\bibfnamefont
  {C.}~\bibnamefont {Stoeckl}}, \bibinfo {author} {\bibfnamefont
  {B.}~\bibnamefont {Yaakobi}}, \bibinfo {author} {\bibfnamefont {J.~A.}\
  \bibnamefont {Frenje}}, \bibinfo {author} {\bibfnamefont {C.~K.}\
  \bibnamefont {Li}}, \bibinfo {author} {\bibfnamefont {R.~D.}\ \bibnamefont
  {Petrasso}}, \ and\ \bibinfo {author} {\bibfnamefont {F.~H.}\ \bibnamefont
  {S\'eguin}},\ }\href@noop {} {\bibfield  {journal} {\bibinfo  {journal}
  {Phys. Rev. Lett.}\ }\textbf {\bibinfo {volume} {100}},\ \bibinfo {pages}
  {185005} (\bibinfo {year} {2008})}\BibitemShut {NoStop}%
\bibitem [{\citenamefont {Sangster}\ \emph {et~al.}(2008)\citenamefont
  {Sangster}, \citenamefont {Goncharov}, \citenamefont {Radha}, \citenamefont
  {Smalyuk}, \citenamefont {Betti}, \citenamefont {Craxton}, \citenamefont
  {Delettrez}, \citenamefont {Edgell}, \citenamefont {Glebov}, \citenamefont
  {Harding}, \citenamefont {Jacobs-Perkins}, \citenamefont {Knauer},
  \citenamefont {Marshall}, \citenamefont {McCrory}, \citenamefont {McKenty},
  \citenamefont {Meyerhofer}, \citenamefont {Regan}, \citenamefont {Seka},
  \citenamefont {Short}, \citenamefont {Skupsky}, \citenamefont {Soures},
  \citenamefont {Stoeckl}, \citenamefont {Yaakobi}, \citenamefont {Shvarts},
  \citenamefont {Frenje}, \citenamefont {Li}, \citenamefont {Petrasso},\ and\
  \citenamefont {S\'eguin}}]{Sangster2008High}%
  \BibitemOpen
  \bibfield  {author} {\bibinfo {author} {\bibfnamefont {T.~C.}\ \bibnamefont
  {Sangster}}, \bibinfo {author} {\bibfnamefont {V.~N.}\ \bibnamefont
  {Goncharov}}, \bibinfo {author} {\bibfnamefont {P.~B.}\ \bibnamefont
  {Radha}}, \bibinfo {author} {\bibfnamefont {V.~A.}\ \bibnamefont {Smalyuk}},
  \bibinfo {author} {\bibfnamefont {R.}~\bibnamefont {Betti}}, \bibinfo
  {author} {\bibfnamefont {R.~S.}\ \bibnamefont {Craxton}}, \bibinfo {author}
  {\bibfnamefont {J.~A.}\ \bibnamefont {Delettrez}}, \bibinfo {author}
  {\bibfnamefont {D.~H.}\ \bibnamefont {Edgell}}, \bibinfo {author}
  {\bibfnamefont {V.~Y.}\ \bibnamefont {Glebov}}, \bibinfo {author}
  {\bibfnamefont {D.~R.}\ \bibnamefont {Harding}}, \bibinfo {author}
  {\bibfnamefont {D.}~\bibnamefont {Jacobs-Perkins}}, \bibinfo {author}
  {\bibfnamefont {J.~P.}\ \bibnamefont {Knauer}}, \bibinfo {author}
  {\bibfnamefont {F.~J.}\ \bibnamefont {Marshall}}, \bibinfo {author}
  {\bibfnamefont {R.~L.}\ \bibnamefont {McCrory}}, \bibinfo {author}
  {\bibfnamefont {P.~W.}\ \bibnamefont {McKenty}}, \bibinfo {author}
  {\bibfnamefont {D.~D.}\ \bibnamefont {Meyerhofer}}, \bibinfo {author}
  {\bibfnamefont {S.~P.}\ \bibnamefont {Regan}}, \bibinfo {author}
  {\bibfnamefont {W.}~\bibnamefont {Seka}}, \bibinfo {author} {\bibfnamefont
  {R.~W.}\ \bibnamefont {Short}}, \bibinfo {author} {\bibfnamefont
  {S.}~\bibnamefont {Skupsky}}, \bibinfo {author} {\bibfnamefont {J.~M.}\
  \bibnamefont {Soures}}, \bibinfo {author} {\bibfnamefont {C.}~\bibnamefont
  {Stoeckl}}, \bibinfo {author} {\bibfnamefont {B.}~\bibnamefont {Yaakobi}},
  \bibinfo {author} {\bibfnamefont {D.}~\bibnamefont {Shvarts}}, \bibinfo
  {author} {\bibfnamefont {J.~A.}\ \bibnamefont {Frenje}}, \bibinfo {author}
  {\bibfnamefont {C.~K.}\ \bibnamefont {Li}}, \bibinfo {author} {\bibfnamefont
  {R.~D.}\ \bibnamefont {Petrasso}}, \ and\ \bibinfo {author} {\bibfnamefont
  {F.~H.}\ \bibnamefont {S\'eguin}},\ }\href@noop {} {\bibfield  {journal}
  {\bibinfo  {journal} {Phys. Rev. Lett.}\ }\textbf {\bibinfo {volume} {100}},\
  \bibinfo {pages} {185006} (\bibinfo {year} {2008})}\BibitemShut {NoStop}%
\bibitem [{\citenamefont {Skupsky}\ \emph {et~al.}(1989)\citenamefont
  {Skupsky}, \citenamefont {Short}, \citenamefont {Kessler}, \citenamefont
  {Craxton}, \citenamefont {Letzring},\ and\ \citenamefont
  {Soures}}]{skupsky1989improved}%
  \BibitemOpen
  \bibfield  {author} {\bibinfo {author} {\bibfnamefont {S.}~\bibnamefont
  {Skupsky}}, \bibinfo {author} {\bibfnamefont {R.}~\bibnamefont {Short}},
  \bibinfo {author} {\bibfnamefont {T.}~\bibnamefont {Kessler}}, \bibinfo
  {author} {\bibfnamefont {R.}~\bibnamefont {Craxton}}, \bibinfo {author}
  {\bibfnamefont {S.}~\bibnamefont {Letzring}}, \ and\ \bibinfo {author}
  {\bibfnamefont {J.}~\bibnamefont {Soures}},\ }\href@noop {} {\bibfield
  {journal} {\bibinfo  {journal} {J. Appl. Phys.}\ }\textbf {\bibinfo {volume}
  {66}},\ \bibinfo {pages} {3456} (\bibinfo {year} {1989})}\BibitemShut
  {NoStop}%
\bibitem [{\citenamefont {Lehmberg}\ and\ \citenamefont
  {Obenschain}(1983)}]{lehmberg1983use}%
  \BibitemOpen
  \bibfield  {author} {\bibinfo {author} {\bibfnamefont {R.}~\bibnamefont
  {Lehmberg}}\ and\ \bibinfo {author} {\bibfnamefont {S.}~\bibnamefont
  {Obenschain}},\ }\href@noop {} {\bibfield  {journal} {\bibinfo  {journal}
  {Opt. Commun.}\ }\textbf {\bibinfo {volume} {46}},\ \bibinfo {pages} {27}
  (\bibinfo {year} {1983})}\BibitemShut {NoStop}%
\bibitem [{\citenamefont {Froula}\ \emph {et~al.}(2010)\citenamefont {Froula},
  \citenamefont {Divol}, \citenamefont {London}, \citenamefont {Berger},
  \citenamefont {D{\"o}ppner}, \citenamefont {Meezan}, \citenamefont {Ralph},
  \citenamefont {Ross}, \citenamefont {Suter},\ and\ \citenamefont
  {Glenzer}}]{froula2010experimental}%
  \BibitemOpen
  \bibfield  {author} {\bibinfo {author} {\bibfnamefont {D.}~\bibnamefont
  {Froula}}, \bibinfo {author} {\bibfnamefont {L.}~\bibnamefont {Divol}},
  \bibinfo {author} {\bibfnamefont {R.}~\bibnamefont {London}}, \bibinfo
  {author} {\bibfnamefont {R.}~\bibnamefont {Berger}}, \bibinfo {author}
  {\bibfnamefont {T.}~\bibnamefont {D{\"o}ppner}}, \bibinfo {author}
  {\bibfnamefont {N.}~\bibnamefont {Meezan}}, \bibinfo {author} {\bibfnamefont
  {J.}~\bibnamefont {Ralph}}, \bibinfo {author} {\bibfnamefont
  {J.}~\bibnamefont {Ross}}, \bibinfo {author} {\bibfnamefont {L.}~\bibnamefont
  {Suter}}, \ and\ \bibinfo {author} {\bibfnamefont {S.}~\bibnamefont
  {Glenzer}},\ }\href@noop {} {\bibfield  {journal} {\bibinfo  {journal} {Phys.
  Plasmas}\ }\textbf {\bibinfo {volume} {17}},\ \bibinfo {pages} {056302}
  (\bibinfo {year} {2010})}\BibitemShut {NoStop}%
\bibitem [{\citenamefont {Moody}\ \emph {et~al.}(2001)\citenamefont {Moody},
  \citenamefont {MacGowan}, \citenamefont {Rothenberg}, \citenamefont {Berger},
  \citenamefont {Divol}, \citenamefont {Glenzer}, \citenamefont {Kirkwood},
  \citenamefont {Williams},\ and\ \citenamefont
  {Young}}]{moody2001backscatter}%
  \BibitemOpen
  \bibfield  {author} {\bibinfo {author} {\bibfnamefont {J.~D.}\ \bibnamefont
  {Moody}}, \bibinfo {author} {\bibfnamefont {B.~J.}\ \bibnamefont {MacGowan}},
  \bibinfo {author} {\bibfnamefont {J.~E.}\ \bibnamefont {Rothenberg}},
  \bibinfo {author} {\bibfnamefont {R.~L.}\ \bibnamefont {Berger}}, \bibinfo
  {author} {\bibfnamefont {L.}~\bibnamefont {Divol}}, \bibinfo {author}
  {\bibfnamefont {S.~H.}\ \bibnamefont {Glenzer}}, \bibinfo {author}
  {\bibfnamefont {R.~K.}\ \bibnamefont {Kirkwood}}, \bibinfo {author}
  {\bibfnamefont {E.~A.}\ \bibnamefont {Williams}}, \ and\ \bibinfo {author}
  {\bibfnamefont {P.~E.}\ \bibnamefont {Young}},\ }\href@noop {} {\bibfield
  {journal} {\bibinfo  {journal} {Phys. Rev. Lett.}\ }\textbf {\bibinfo
  {volume} {86}},\ \bibinfo {pages} {2810} (\bibinfo {year}
  {2001})}\BibitemShut {NoStop}%
\bibitem [{\citenamefont {Albright}\ \emph {et~al.}(2014)\citenamefont
  {Albright}, \citenamefont {Yin},\ and\ \citenamefont
  {Afeyan}}]{B2014Control}%
  \BibitemOpen
  \bibfield  {author} {\bibinfo {author} {\bibfnamefont {B.~J.}\ \bibnamefont
  {Albright}}, \bibinfo {author} {\bibfnamefont {L.}~\bibnamefont {Yin}}, \
  and\ \bibinfo {author} {\bibfnamefont {B.}~\bibnamefont {Afeyan}},\
  }\href@noop {} {\bibfield  {journal} {\bibinfo  {journal} {Phys.rev.lett}\
  }\textbf {\bibinfo {volume} {113}},\ \bibinfo {pages} {045002} (\bibinfo
  {year} {2014})}\BibitemShut {NoStop}%
\bibitem [{\citenamefont {Thomson}\ and\ \citenamefont
  {Karush}(1974)}]{thomson1974effects}%
  \BibitemOpen
  \bibfield  {author} {\bibinfo {author} {\bibfnamefont {J.}~\bibnamefont
  {Thomson}}\ and\ \bibinfo {author} {\bibfnamefont {J.~I.}\ \bibnamefont
  {Karush}},\ }\href@noop {} {\bibfield  {journal} {\bibinfo  {journal} {Phys.
  Fluids}\ }\textbf {\bibinfo {volume} {17}},\ \bibinfo {pages} {1608}
  (\bibinfo {year} {1974})}\BibitemShut {NoStop}%
\bibitem [{\citenamefont {Eimerl}\ \emph {et~al.}(1992)\citenamefont {Eimerl},
  \citenamefont {Kruer},\ and\ \citenamefont {Campbell}}]{Eimerl}%
  \BibitemOpen
  \bibfield  {author} {\bibinfo {author} {\bibfnamefont {D.}~\bibnamefont
  {Eimerl}}, \bibinfo {author} {\bibfnamefont {W.~L.}\ \bibnamefont {Kruer}}, \
  and\ \bibinfo {author} {\bibfnamefont {E.~M.}\ \bibnamefont {Campbell}},\
  }\href@noop {} {\bibfield  {journal} {\bibinfo  {journal} {Comments Plasma
  Phys. Controlled Fusion}\ }\textbf {\bibinfo {volume} {15}},\ \bibinfo
  {pages} {85} (\bibinfo {year} {1992})}\BibitemShut {NoStop}%
\bibitem [{\citenamefont {Turner}\ \emph {et~al.}(1985)\citenamefont {Turner},
  \citenamefont {Estabrook}, \citenamefont {Kauffman}, \citenamefont {Bach},
  \citenamefont {Drake}, \citenamefont {Phillion}, \citenamefont {Lasinski},
  \citenamefont {Campbell}, \citenamefont {Kruer},\ and\ \citenamefont
  {Williams}}]{turner1985evidence}%
  \BibitemOpen
  \bibfield  {author} {\bibinfo {author} {\bibfnamefont {R.~E.}\ \bibnamefont
  {Turner}}, \bibinfo {author} {\bibfnamefont {K.}~\bibnamefont {Estabrook}},
  \bibinfo {author} {\bibfnamefont {R.~L.}\ \bibnamefont {Kauffman}}, \bibinfo
  {author} {\bibfnamefont {D.~R.}\ \bibnamefont {Bach}}, \bibinfo {author}
  {\bibfnamefont {R.~P.}\ \bibnamefont {Drake}}, \bibinfo {author}
  {\bibfnamefont {D.~W.}\ \bibnamefont {Phillion}}, \bibinfo {author}
  {\bibfnamefont {B.~F.}\ \bibnamefont {Lasinski}}, \bibinfo {author}
  {\bibfnamefont {E.~M.}\ \bibnamefont {Campbell}}, \bibinfo {author}
  {\bibfnamefont {W.~L.}\ \bibnamefont {Kruer}}, \ and\ \bibinfo {author}
  {\bibfnamefont {E.~A.}\ \bibnamefont {Williams}},\ }\href@noop {} {\bibfield
  {journal} {\bibinfo  {journal} {Phys. Rev. Lett.}\ }\textbf {\bibinfo
  {volume} {54}},\ \bibinfo {pages} {189} (\bibinfo {year} {1985})}\BibitemShut
  {NoStop}%
\bibitem [{\citenamefont {Craxton}\ \emph {et~al.}(2015)\citenamefont
  {Craxton}, \citenamefont {Anderson}, \citenamefont {Boehly}, \citenamefont
  {Goncharov}, \citenamefont {Harding}, \citenamefont {Knauer}, \citenamefont
  {McCrory}, \citenamefont {McKenty}, \citenamefont {Meyerhofer}, \citenamefont
  {Myatt} \emph {et~al.}}]{craxton2015direct}%
  \BibitemOpen
  \bibfield  {author} {\bibinfo {author} {\bibfnamefont {R.~S.}\ \bibnamefont
  {Craxton}}, \bibinfo {author} {\bibfnamefont {K.~S.}\ \bibnamefont
  {Anderson}}, \bibinfo {author} {\bibfnamefont {T.~R.}\ \bibnamefont
  {Boehly}}, \bibinfo {author} {\bibfnamefont {V.~N.}\ \bibnamefont
  {Goncharov}}, \bibinfo {author} {\bibfnamefont {D.~R.}\ \bibnamefont
  {Harding}}, \bibinfo {author} {\bibfnamefont {J.~P.}\ \bibnamefont {Knauer}},
  \bibinfo {author} {\bibfnamefont {R.~L.}\ \bibnamefont {McCrory}}, \bibinfo
  {author} {\bibfnamefont {P.~W.}\ \bibnamefont {McKenty}}, \bibinfo {author}
  {\bibfnamefont {D.~D.}\ \bibnamefont {Meyerhofer}}, \bibinfo {author}
  {\bibfnamefont {J.~F.}\ \bibnamefont {Myatt}},  \emph {et~al.},\ }\href@noop
  {} {\bibfield  {journal} {\bibinfo  {journal} {Phys. Plasmas}\ }\textbf
  {\bibinfo {volume} {22}},\ \bibinfo {pages} {110501} (\bibinfo {year}
  {2015})}\BibitemShut {NoStop}%
\bibitem [{\citenamefont {Strozzi}\ \emph {et~al.}(2017)\citenamefont
  {Strozzi}, \citenamefont {Bailey}, \citenamefont {Michel}, \citenamefont
  {Divol}, \citenamefont {Sepke}, \citenamefont {Kerbel}, \citenamefont
  {Thomas}, \citenamefont {Ralph}, \citenamefont {Moody},\ and\ \citenamefont
  {Schneider}}]{InterplayStrozzi}%
  \BibitemOpen
  \bibfield  {author} {\bibinfo {author} {\bibfnamefont {D.~J.}\ \bibnamefont
  {Strozzi}}, \bibinfo {author} {\bibfnamefont {D.~S.}\ \bibnamefont {Bailey}},
  \bibinfo {author} {\bibfnamefont {P.}~\bibnamefont {Michel}}, \bibinfo
  {author} {\bibfnamefont {L.}~\bibnamefont {Divol}}, \bibinfo {author}
  {\bibfnamefont {S.~M.}\ \bibnamefont {Sepke}}, \bibinfo {author}
  {\bibfnamefont {G.~D.}\ \bibnamefont {Kerbel}}, \bibinfo {author}
  {\bibfnamefont {C.~A.}\ \bibnamefont {Thomas}}, \bibinfo {author}
  {\bibfnamefont {J.~E.}\ \bibnamefont {Ralph}}, \bibinfo {author}
  {\bibfnamefont {J.~D.}\ \bibnamefont {Moody}}, \ and\ \bibinfo {author}
  {\bibfnamefont {M.~B.}\ \bibnamefont {Schneider}},\ }\href@noop {} {\bibfield
   {journal} {\bibinfo  {journal} {Phys. Rev. Lett.}\ }\textbf {\bibinfo
  {volume} {118}},\ \bibinfo {pages} {025002} (\bibinfo {year}
  {2017})}\BibitemShut {NoStop}%
\bibitem [{\citenamefont {Mourou}\ \emph {et~al.}(2013)\citenamefont {Mourou},
  \citenamefont {Brocklesby}, \citenamefont {Tajima},\ and\ \citenamefont
  {Limpert}}]{mourou2013future}%
  \BibitemOpen
  \bibfield  {author} {\bibinfo {author} {\bibfnamefont {G.}~\bibnamefont
  {Mourou}}, \bibinfo {author} {\bibfnamefont {B.}~\bibnamefont {Brocklesby}},
  \bibinfo {author} {\bibfnamefont {T.}~\bibnamefont {Tajima}}, \ and\ \bibinfo
  {author} {\bibfnamefont {J.}~\bibnamefont {Limpert}},\ }\href@noop {}
  {\bibfield  {journal} {\bibinfo  {journal} {Nature Photon.}\ }\textbf
  {\bibinfo {volume} {7}},\ \bibinfo {pages} {258} (\bibinfo {year}
  {2013})}\BibitemShut {NoStop}%
\bibitem [{\citenamefont {Andrusyak}\ \emph {et~al.}(2009)\citenamefont
  {Andrusyak}, \citenamefont {Smirnov}, \citenamefont {Venus},\ and\
  \citenamefont {Glebov}}]{andrusyak2009beam}%
  \BibitemOpen
  \bibfield  {author} {\bibinfo {author} {\bibfnamefont {O.}~\bibnamefont
  {Andrusyak}}, \bibinfo {author} {\bibfnamefont {V.}~\bibnamefont {Smirnov}},
  \bibinfo {author} {\bibfnamefont {G.}~\bibnamefont {Venus}}, \ and\ \bibinfo
  {author} {\bibfnamefont {L.}~\bibnamefont {Glebov}},\ }\href@noop {}
  {\bibfield  {journal} {\bibinfo  {journal} {Opt. Commun.}\ }\textbf {\bibinfo
  {volume} {282}},\ \bibinfo {pages} {2560} (\bibinfo {year}
  {2009})}\BibitemShut {NoStop}%
\bibitem [{\citenamefont {Hamilton}\ \emph {et~al.}(2004)\citenamefont
  {Hamilton}, \citenamefont {Tidwell}, \citenamefont {Meekhof}, \citenamefont
  {Seamans}, \citenamefont {Gitkind},\ and\ \citenamefont
  {Lowenthal}}]{hamilton2004high}%
  \BibitemOpen
  \bibfield  {author} {\bibinfo {author} {\bibfnamefont {C.~E.}\ \bibnamefont
  {Hamilton}}, \bibinfo {author} {\bibfnamefont {S.~C.}\ \bibnamefont
  {Tidwell}}, \bibinfo {author} {\bibfnamefont {D.}~\bibnamefont {Meekhof}},
  \bibinfo {author} {\bibfnamefont {J.}~\bibnamefont {Seamans}}, \bibinfo
  {author} {\bibfnamefont {N.}~\bibnamefont {Gitkind}}, \ and\ \bibinfo
  {author} {\bibfnamefont {D.~D.}\ \bibnamefont {Lowenthal}},\ }in\ \href@noop
  {} {\emph {\bibinfo {booktitle} {Lasers and Applications in Science and
  Engineering}}}\ (\bibinfo {organization} {International Society for Optics
  and Photonics},\ \bibinfo {year} {2004})\ pp.\ \bibinfo {pages}
  {1--10}\BibitemShut {NoStop}%
\bibitem [{\citenamefont {Farmer}\ \emph {et~al.}(1999)\citenamefont {Farmer},
  \citenamefont {Lowenthal},\ and\ \citenamefont
  {Pierce}}]{farmer1999incoherent}%
  \BibitemOpen
  \bibfield  {author} {\bibinfo {author} {\bibfnamefont {J.~N.}\ \bibnamefont
  {Farmer}}, \bibinfo {author} {\bibfnamefont {D.}~\bibnamefont {Lowenthal}}, \
  and\ \bibinfo {author} {\bibfnamefont {J.}~\bibnamefont {Pierce}},\ }in\
  \href@noop {} {\emph {\bibinfo {booktitle} {LEOS'99. IEEE Lasers and
  Electro-Optics Society 1999 12th Annual Meeting}}},\ Vol.~\bibinfo {volume}
  {2}\ (\bibinfo {organization} {IEEE},\ \bibinfo {year} {1999})\ pp.\ \bibinfo
  {pages} {796--797}\BibitemShut {NoStop}%
\bibitem [{\citenamefont {Benedetti}\ \emph {et~al.}(2014)\citenamefont
  {Benedetti}, \citenamefont {Schroeder}, \citenamefont {Esarey},\ and\
  \citenamefont {Leemans}}]{benedetti2014plasma}%
  \BibitemOpen
  \bibfield  {author} {\bibinfo {author} {\bibfnamefont {C.}~\bibnamefont
  {Benedetti}}, \bibinfo {author} {\bibfnamefont {C.}~\bibnamefont
  {Schroeder}}, \bibinfo {author} {\bibfnamefont {E.}~\bibnamefont {Esarey}}, \
  and\ \bibinfo {author} {\bibfnamefont {W.}~\bibnamefont {Leemans}},\
  }\href@noop {} {\bibfield  {journal} {\bibinfo  {journal} {Phys. Plasmas}\
  }\textbf {\bibinfo {volume} {21}},\ \bibinfo {pages} {056706} (\bibinfo
  {year} {2014})}\BibitemShut {NoStop}%
\bibitem [{\citenamefont {Eimerl}\ \emph {et~al.}(2016)\citenamefont {Eimerl},
  \citenamefont {Skupsky}, \citenamefont {Myatt},\ and\ \citenamefont
  {Campbell}}]{eimerl2016stardriver}%
  \BibitemOpen
  \bibfield  {author} {\bibinfo {author} {\bibfnamefont {D.}~\bibnamefont
  {Eimerl}}, \bibinfo {author} {\bibfnamefont {S.}~\bibnamefont {Skupsky}},
  \bibinfo {author} {\bibfnamefont {J.}~\bibnamefont {Myatt}}, \ and\ \bibinfo
  {author} {\bibfnamefont {E.~M.}\ \bibnamefont {Campbell}},\ }\href@noop {}
  {\bibfield  {journal} {\bibinfo  {journal} {J. Fusion Energy}\ }\textbf
  {\bibinfo {volume} {35}},\ \bibinfo {pages} {459} (\bibinfo {year}
  {2016})}\BibitemShut {NoStop}%
\bibitem [{\citenamefont {Kruer}(1988)}]{kruer1988physics}%
  \BibitemOpen
  \bibfield  {author} {\bibinfo {author} {\bibfnamefont {W.~L.}\ \bibnamefont
  {Kruer}},\ }\href@noop {} {\emph {\bibinfo {title} {The Physics of Laser
  Plasma Interactions}}},\ Vol.~\bibinfo {volume} {70}\ (\bibinfo  {publisher}
  {Addison-Wesley New York},\ \bibinfo {year} {1988})\BibitemShut {NoStop}%
\bibitem [{\citenamefont {Zhao}\ \emph {et~al.}(2014)\citenamefont {Zhao},
  \citenamefont {Zheng}, \citenamefont {Chen}, \citenamefont {Yu},
  \citenamefont {Weng}, \citenamefont {Ren}, \citenamefont {Liu},\ and\
  \citenamefont {Sheng}}]{zhao2014effects}%
  \BibitemOpen
  \bibfield  {author} {\bibinfo {author} {\bibfnamefont {Y.}~\bibnamefont
  {Zhao}}, \bibinfo {author} {\bibfnamefont {J.}~\bibnamefont {Zheng}},
  \bibinfo {author} {\bibfnamefont {M.}~\bibnamefont {Chen}}, \bibinfo {author}
  {\bibfnamefont {L.~L.}\ \bibnamefont {Yu}}, \bibinfo {author} {\bibfnamefont
  {S.~M.}\ \bibnamefont {Weng}}, \bibinfo {author} {\bibfnamefont
  {C.}~\bibnamefont {Ren}}, \bibinfo {author} {\bibfnamefont {C.~S.}\
  \bibnamefont {Liu}}, \ and\ \bibinfo {author} {\bibfnamefont {Z.~M.}\
  \bibnamefont {Sheng}},\ }\href@noop {} {\bibfield  {journal} {\bibinfo
  {journal} {Phys. Plasmas}\ }\textbf {\bibinfo {volume} {21}},\ \bibinfo
  {pages} {112114} (\bibinfo {year} {2014})}\BibitemShut {NoStop}%
\bibitem [{\citenamefont {Chen}\ \emph {et~al.}(2008)\citenamefont {Chen},
  \citenamefont {Sheng}, \citenamefont {Zheng}, \citenamefont {Ma},\ and\
  \citenamefont {Zhang}}]{chen2008development}%
  \BibitemOpen
  \bibfield  {author} {\bibinfo {author} {\bibfnamefont {M.}~\bibnamefont
  {Chen}}, \bibinfo {author} {\bibfnamefont {Z.~M.}\ \bibnamefont {Sheng}},
  \bibinfo {author} {\bibfnamefont {J.}~\bibnamefont {Zheng}}, \bibinfo
  {author} {\bibfnamefont {Y.~Y.}\ \bibnamefont {Ma}}, \ and\ \bibinfo {author}
  {\bibfnamefont {J.}~\bibnamefont {Zhang}},\ }\href@noop {} {\bibfield
  {journal} {\bibinfo  {journal} {Chin. J. Comput. Phys.}\ }\textbf {\bibinfo
  {volume} {25}},\ \bibinfo {pages} {43} (\bibinfo {year} {2008})}\BibitemShut
  {NoStop}%
\bibitem [{\citenamefont {Myatt}\ \emph {et~al.}(2014)\citenamefont {Myatt},
  \citenamefont {Zhang}, \citenamefont {Short}, \citenamefont {Maximov},
  \citenamefont {Seka}, \citenamefont {Froula}, \citenamefont {Edgell},
  \citenamefont {Michel}, \citenamefont {Igumenshchev}, \citenamefont {Hinkel}
  \emph {et~al.}}]{myatt2014multiple}%
  \BibitemOpen
  \bibfield  {author} {\bibinfo {author} {\bibfnamefont {J.}~\bibnamefont
  {Myatt}}, \bibinfo {author} {\bibfnamefont {J.}~\bibnamefont {Zhang}},
  \bibinfo {author} {\bibfnamefont {R.}~\bibnamefont {Short}}, \bibinfo
  {author} {\bibfnamefont {A.}~\bibnamefont {Maximov}}, \bibinfo {author}
  {\bibfnamefont {W.}~\bibnamefont {Seka}}, \bibinfo {author} {\bibfnamefont
  {D.}~\bibnamefont {Froula}}, \bibinfo {author} {\bibfnamefont
  {D.}~\bibnamefont {Edgell}}, \bibinfo {author} {\bibfnamefont
  {D.}~\bibnamefont {Michel}}, \bibinfo {author} {\bibfnamefont
  {I.}~\bibnamefont {Igumenshchev}}, \bibinfo {author} {\bibfnamefont
  {D.}~\bibnamefont {Hinkel}},  \emph {et~al.},\ }\href@noop {} {\bibfield
  {journal} {\bibinfo  {journal} {Phys. Plasmas}\ }\textbf {\bibinfo {volume}
  {21}},\ \bibinfo {pages} {055501} (\bibinfo {year} {2014})}\BibitemShut
  {NoStop}%
\bibitem [{\citenamefont {Liu}\ \emph {et~al.}(1974)\citenamefont {Liu},
  \citenamefont {Rosenbluth},\ and\ \citenamefont {White}}]{liu1974raman}%
  \BibitemOpen
  \bibfield  {author} {\bibinfo {author} {\bibfnamefont {C.}~\bibnamefont
  {Liu}}, \bibinfo {author} {\bibfnamefont {M.~N.}\ \bibnamefont {Rosenbluth}},
  \ and\ \bibinfo {author} {\bibfnamefont {R.~B.}\ \bibnamefont {White}},\
  }\href@noop {} {\bibfield  {journal} {\bibinfo  {journal} {Phys. Fluids}\
  }\textbf {\bibinfo {volume} {17}},\ \bibinfo {pages} {1211} (\bibinfo {year}
  {1974})}\BibitemShut {NoStop}%
\bibitem [{\citenamefont {Zhao}\ \emph {et~al.}(2015)\citenamefont {Zhao},
  \citenamefont {Yu}, \citenamefont {Zheng}, \citenamefont {Weng},
  \citenamefont {Ren}, \citenamefont {Liu},\ and\ \citenamefont
  {Sheng}}]{zhao2015effects}%
  \BibitemOpen
  \bibfield  {author} {\bibinfo {author} {\bibfnamefont {Y.}~\bibnamefont
  {Zhao}}, \bibinfo {author} {\bibfnamefont {L.~L.}\ \bibnamefont {Yu}},
  \bibinfo {author} {\bibfnamefont {J.}~\bibnamefont {Zheng}}, \bibinfo
  {author} {\bibfnamefont {S.~M.}\ \bibnamefont {Weng}}, \bibinfo {author}
  {\bibfnamefont {C.}~\bibnamefont {Ren}}, \bibinfo {author} {\bibfnamefont
  {C.~S.}\ \bibnamefont {Liu}}, \ and\ \bibinfo {author} {\bibfnamefont
  {Z.~M.}\ \bibnamefont {Sheng}},\ }\href@noop {} {\bibfield  {journal}
  {\bibinfo  {journal} {Phys. Plasmas}\ }\textbf {\bibinfo {volume} {22}},\
  \bibinfo {pages} {052119} (\bibinfo {year} {2015})}\BibitemShut {NoStop}%
\bibitem [{\citenamefont {Ho}\ and\ \citenamefont {Kahn}(1993)}]{Ho1993}%
  \BibitemOpen
  \bibfield  {author} {\bibinfo {author} {\bibfnamefont {K.}~\bibnamefont
  {Ho}}\ and\ \bibinfo {author} {\bibfnamefont {J.}~\bibnamefont {Kahn}},\
  }\href@noop {} {\bibfield  {journal} {\bibinfo  {journal} {IEEE Photon. Tech.
  Lett.}\ }\textbf {\bibinfo {volume} {5}},\ \bibinfo {pages} {721} (\bibinfo
  {year} {1993})}\BibitemShut {NoStop}%
\bibitem [{\citenamefont {Yu}\ \emph {et~al.}(2016)\citenamefont {Yu},
  \citenamefont {Zhao}, \citenamefont {Qian}, \citenamefont {Chen},
  \citenamefont {Weng}, \citenamefont {Sheng}, \citenamefont {Jaroszynski},
  \citenamefont {Mori},\ and\ \citenamefont {Zhang}}]{Yu2016}%
  \BibitemOpen
  \bibfield  {author} {\bibinfo {author} {\bibfnamefont {L.~L.}\ \bibnamefont
  {Yu}}, \bibinfo {author} {\bibfnamefont {Y.}~\bibnamefont {Zhao}}, \bibinfo
  {author} {\bibfnamefont {L.~J.}\ \bibnamefont {Qian}}, \bibinfo {author}
  {\bibfnamefont {M.}~\bibnamefont {Chen}}, \bibinfo {author} {\bibfnamefont
  {S.~M.}\ \bibnamefont {Weng}}, \bibinfo {author} {\bibfnamefont {Z.-M.}\
  \bibnamefont {Sheng}}, \bibinfo {author} {\bibfnamefont {D.}~\bibnamefont
  {Jaroszynski}}, \bibinfo {author} {\bibfnamefont {W.}~\bibnamefont {Mori}}, \
  and\ \bibinfo {author} {\bibfnamefont {J.}~\bibnamefont {Zhang}},\
  }\href@noop {} {\bibfield  {journal} {\bibinfo  {journal} {Nature Commun.}\
  }\textbf {\bibinfo {volume} {7}},\ \bibinfo {pages} {11893} (\bibinfo {year}
  {2016})}\BibitemShut {NoStop}%
\bibitem [{\citenamefont {Dabu}(2010)}]{Dabu2010}%
  \BibitemOpen
  \bibfield  {author} {\bibinfo {author} {\bibfnamefont {R.}~\bibnamefont
  {Dabu}},\ }\href@noop {} {\bibfield  {journal} {\bibinfo  {journal} {Opt.
  Express}\ }\textbf {\bibinfo {volume} {18}},\ \bibinfo {pages} {11689}
  (\bibinfo {year} {2010})}\BibitemShut {NoStop}%
\end{thebibliography}

%

\end{document}